\documentclass[preprintnumbers,eqsecnum,showpacs,aps,prd,epsf,nofootinbib]{revtex4}
\usepackage{latexsym,amsmath}
\usepackage{amssymb} 
\usepackage{color}
\usepackage{graphicx,bm}
\usepackage{dcolumn}    
\usepackage{subfigure}  

\newcommand{\gsim}{\mbox{\raisebox{-1.ex}{$\stackrel
      {\textstyle>}{\textstyle\sim}$}}}
\newcommand{\lsim}{\mbox{\raisebox{-1.ex}{$\stackrel
      {\textstyle<}{\textstyle \sim}$}}}

\def\til#1{\tilde{#1}}

\begin{document}
\thispagestyle{empty}

\title{Thermal Equilibrium of String Gas in Hagedorn Universe}

\author{Yu-ichi Takamizu$^{1}$}
\email{takamizu_at_gravity.phys.waseda.ac.jp}

\author{Hideaki Kudoh$^{2,3}$}
\email{kudoh_at_utap.phys.s.u-tokyo.ac.jp}

\address{
\\
$^{1}$ Department of Physics, Waseda University,
Okubo 3-4-1, Shinjuku, Tokyo 169-8555, Japan
\\
$^{2}$ Department of Physics, UCSB, Santa Barbara, CA 93106, USA
\\
$^{3}$ Department of Physics, The University of Tokyo, 
Tokyo 113-0033, Japan
}

\preprint{UTAP-563}

\begin{abstract}
The thermal equilibrium of string gas is necessary to activate the Brandenberger-Vafa mechanism, which makes our observed 4-dimensional universe enlarge. 
Nevertheless, the thermal equilibrium is not realized in the original setup, a problem that remains as a critical defect. 
We study thermal equilibrium in the Hagedorn universe, and explore possibilities for avoiding the issue aforementioned flaw. 
We employ a minimal modification of the original setup, introducing a dilaton potential. 
Two types of potential are investigated: exponential and double-well potentials. 
For the first type, the basic evolutions of universe and dilaton are such that both the radius of the universe and the dilaton asymptotically grow in over a short time, or that the radius converges to a constant value while the dilaton rolls down toward the weak coupling limit. 
For the second type, in addition to the above solutions, there is another solution in which the dilaton is stabilized at a minimum of potential and the radius grows in proportion to $t$. 
Thermal equilibrium is realized for both cases during the initial phase. 
These simple setups provide possible resolutions of the difficulty. 
\end{abstract}

\maketitle

\section{introduction}

String theory represents the most useful and promising candidate for a unified theory of the fundamental interactions including gravity. 
From cosmological aspects, an ultimate goal of such theory is to resolve three interesting problems: the cosmological constant, the initial singularity problem, and the origin of the three spatial dimensions and time.

For the initial singularity problems, the tachyon condensation of winding strings may play a role when the radius of universe shrinks smaller than the string scale, according to a recent proposal~\cite{McGreevy:2005ci}. 
Here, we focus on the dimensionality problem in the context of the Brandenberger-Vafa (BV) scenario \cite{Brandenberger_Vafa} 
(see \cite{Durrer:2005nz,Karch:2005yz}). 
In the original BV scenario \cite{Brandenberger_Vafa}, it is assumed that all nine spatial dimensions start from the toroidally compactified radii near the string length and the universe 
is filled with 
an ideal gas of fundamental matter, so-called string gas. 
It is also assumed that the string gas 
is initially thermalized at the critical temperature $T_H$, 
called the Hagedorn temperature \cite{Hagedorn}. 
In order to resolve the dimensionality problem, 
the string winding modes play a particularly important role. 
The winding strings prevent the dimensions which they wrap 
from expanding, as shown in \cite{Tseytlin:1991xk,Tseytlin:1991ss}. 
The annihilation of winding and anti-winding strings determines how many 
dimensions expand and how many dimensions stay at the string length. 
A simple counting argument suggests that this annihilation occurs mostly 
in the space-time dimensions of $D=4$, so the three spatial dimensions become large. 
Some studies have already examined various aspects of the BV scenario and 
it has been extended in a variety of ways 
\cite {Tseytlin:1991xk,Tseytlin:1991ss,Sakellariadou:1995vk,Easther:2004sd,Alexander:2000xv,Bassett:2003ck,Danos:2004jz,Berndsen:2004tj,Campos:2005da,Brandenberger:2005bd,Borunda:2006fx,Brandenberger:2006pr}. 
(See also recent reviews 
\cite{Brandenberger:2005nz,Brandenberger:2005fb,Battefeld:2005av}).

The interaction (annihilation) rate $\Gamma$ of the string winding modes, 
with the string coupling $e^{\phi}$, is roughly 
\begin{align}
\Gamma\simeq 100 \ln \,E \,e^{4\phi}\,.
\label{interaction_rate}
\end{align}
(See \cite{Danos:2004jz} and Sec.~\ref{sec:thermal equilibrium} for more 
detailed discussion). 
Assuming that this interaction works efficiently at an early stage of the universe, decompactification proceeds. 
However, there are three main assumptions on this scenario: adiabatic evolution, weak coupling and thermal equilibrium.
A critical point of this scenario is the last assumption about thermal equilibrium \cite{Danos:2004jz} (see also \cite{Easther:2004sd} for another assessment).

The thermal equilibrium condition is given by 
\begin{eqnarray}
     \Gamma  > H ,
\end{eqnarray}
where $H$ is the Hubble expansion rate. 
Let us briefly estimate this condition based on a typical cosmological evolution at the Hagedorn temperature with ``string matter". 
The radius of the universe asymptotes to a constant value, whose order is the string scale or several times of it. 
This means that all dimensions will be still small, which is inconsistent with our large four dimensions. 
However, if the radii (of three dimensions) asymptote to a sufficiently large value, then the universe could be matched to a radiation dominated phase. 
This type of evolution is only allowed when the initial dilaton satisfies a certain condition. 
From the condition indicated with Eq. (\ref{interaction_rate}), the interaction rate is found to be bounded from above as 
\begin{eqnarray}
  \Gamma ~\lesssim \Bigl(O(10^{-1}) H_0^3\Bigr)H_0\,,
\label{eq:Gamma < H estimate}
\end{eqnarray}
where the subscript $0$ denotes the value at an initial time 
(Sec. \ref{sec:thermal equilibrium}). 
This shows that the thermal equilibrium condition is not 
satisfied at the initial phase of the universe, under the adiabatic condition $H \lesssim 1$. 
Therefore, no consistent treatment with well-defined thermodynamic functions can be carried out~\cite{Deo:1989bv}, and all analysis based on such an approach cannot be trusted.

This non-equilibrium aspect of string gas is related to cosmological expansion at the Hagedorn phase. 
Some features are necessary for resolving this difficulty, for instance: 
\begin{itemize}
  \item  The universe does not evolve toward a constant radius at late times. 
  \item  Initial evolution of the universe is modified. 
\end{itemize}
In the latter case, 
the universe may or may not evolve toward a constant radius, and 
the initial constraint for the dilation may be replaced by a weaker condition. 
There are several possibilities for achieving the above two evolutions. 
In the straightforward derivation of the interaction rate (\ref{eq:Gamma < H estimate}), we have assumed a simple dilaton-gravity system without taking into account any nontrivial coupling between the dilaton and 
``string matter" (gas) fields. 
Any nontrivial coupling modifies the dynamics of dilaton and cosmological expansion, and it may weaken the difficult requirement for the initial condition. 
For example, incorporating the NS-NS and R-R fields into the action \cite{Campos:2005da,Brandenberger:2005bd}, or the effects of higher curvature corrections \cite{Borunda:2006fx} will alter the evolution of the universe at an early stage.

With these ideas in mind, 
we wish to pursue the possibility of resolving the thermal equilibrium issue. 
We adopt a simple modification of the scenario by taking into account the dilaton potential. 
This simple alternative will allow us to study the system rather extensively and give insight into other possibilities and approaches. 
We will see that the above two features are realized in the simple models in the present paper.

The paper is organized as follows. 
In the next section, some general aspects of dilaton-gravity and string gas in the extreme Hagedorn regime of high-energy densities are briefly reviewed. 
We will take the dilaton potential to be of two types. 
The first case is discussed in Sec. III, the second in Sec IV. 
For both cases, we analyze the dynamics of the Hagedorn regime with a single scale factor (Hubble radius) and with large and small radii.  
In Sec. V we discuss the thermal equilibrium of string gas for the two models. 
The final section is devoted to summary and discussion. 
We adopt the string scale $\alpha'=1$.

\section{String Gas in Dilaton Gravity} 

\subsection{Basic equations}

Dilaton-gravity comes from the low-energy effective action of string theory. Ignoring contributions from the antisymmetric two-form and including a potential $V(\phi)$ for the dilaton, 
\footnote{We give the dilaton potential in the  string frame, while the corresponding potential in the Einstein frame is given in Appendix \ref{effective_potential}. 
Note that the initial data in the Einstein frame also differ from those in the string frame.}
the action of this system is described as 
\begin{eqnarray}
  S 
  =
  \int d^{10} x \sqrt{|g|}\bigl[e^{-2\phi} \left( R+4(\nabla \phi)^2 + V \right)
  + {\cal{L}}_M
  \bigr] \,,
\end{eqnarray}
where $g$ is the determinant of the background metric $g_{\mu\nu}$ 
and ${\cal{L}}_M$ is a Lagrangian of some matter. 
The coupling of $\phi$ with gravity is the standard one arising in string theory. 
Assuming a spatially-homogeneous universe, 
\begin{eqnarray}
&& ds^2  =  -dt^2+\sum_{i=1}^{9}e^{2\lambda_i(t)}dx_i^2, 
\cr
&& \phi=\phi(t)\,,
\end{eqnarray}
we can reduce the action to 
\begin{align}
S=\int dt \sqrt{-g_{00}} (-g^{00}) \bigl[&e^{-\psi}(\sum_{i=1}^9 {\dot{\lambda}_i}^2-\dot{\psi}^2+V(\phi))-F(\lambda_i, \beta \sqrt{-g_{00}}) \bigr]\,,
\end{align}
where we have introduced a shifted dilaton $\psi$, 
\begin{equation}
 \psi\equiv 2\phi-\sum_{i=1}^9\lambda_i\,,
\end{equation}
to simplify the obtained equation of motion.
In the reduced action, we have introduced the (one loop) free energy $F$ of a closed string as the Lagrangian of matter. 
This is only possible in an early universe in which the string gas is in thermal equilibrium at the temperature $\beta^{-1}$. 
Variation with respect to $g_{00}$, $\lambda_i$, $\psi$ yields the following equations of motion: 
\begin{eqnarray}
   &&-\sum_{i=1}^9{\dot{\lambda}_i}^2+\dot{\psi}^2=e^{\psi}E-V(\phi)\,,  
\cr
   &&\ddot{\lambda}_i-\dot{\psi}\dot{\lambda}_i={1\over 2}e^{\psi}P_i+{1\over 4}
   V'(\phi)\,,
\label{eq:eoms1}
\\
   && \ddot{\psi}-{1\over 2}\sum_{i=1}^9{\dot{\lambda}_i}^2-{1\over 2}\dot{\psi}^2
    ={1\over 2}V(\phi)-{1\over 4}V'(\phi)\,,
\nonumber
\end{eqnarray}
where $E=\rho e^{\sum \lambda_i}$ is the total energy and $P_i=p_i e^{\sum \lambda_i}$ is the total pressure along the 
$i$-th direction, respectively, obtained by multiplying the total spatial volume $e^{\sum \lambda_i}$ with the energy density $\rho$ and pressure $p_i$. 
These quantities are related to the free energy $F$ by the basic thermodynamic relation: 
\begin{align}
 & E = F+\beta\frac{\partial F}{\partial \beta}\,,
\cr
 & P_i=-\frac{\partial F}{\partial \lambda_i}\,.
 \nonumber
\end{align}
Employing Eqs. (\ref{eq:eoms1}), the conservation of the total energy is 
\begin{equation}
  \dot{E} + \sum_i^9 \dot{\lambda}_i P_i=0\,,
  \label{conserve_E}
\end{equation}
or equivalently, 
\begin{equation}
  \dot{\rho}+\sum_i^9 \dot{\lambda}_i (\rho + p_i)=0\,.
  \label{conserve_rho}
\end{equation}
The basic equations become simple forms in terms of the shifted dilaton. 
In some cases, however, the equations written by the original dilaton are convenient.

Here, if we separate the spatial dimensions into $d$-spatial large dimensions and 
$(9-d)$-spatial small dimensions, which are denoted as 
\begin{equation}
  R = e^{\mu}, ~~r=e^{\nu}\,,
\end{equation}
respectively,  Eqs. (\ref{eq:eoms1}) in terms of the original dilaton take the following forms: 
\begin{align}
&  - d\dot{\mu}^2-(9-d)\dot{\nu}^2+(2\dot{\phi}-d\dot{\mu}-(9-d)\dot{\nu})^2
= e^{2\phi}\rho-V(\phi)  \,,
\nonumber
\\
&  \ddot{\mu}-(2\dot{\phi}-d\, \dot{\mu}-(9-d)\dot{\nu})
  \dot{\mu}={1\over 2}e^{2\phi}
  p_{d}+{1\over 4} V'(\phi)\,,
\nonumber
\\
& 
  \ddot{\nu}-(2\dot{\phi}-d\, \dot{\mu}-(9-d) \dot{\nu})\dot{\nu}
  = {1\over 2}e^{2\phi}
  p_{~9-d}+{1\over 4} V'(\phi)\,,
\nonumber
\\
& 
  \ddot{\phi} - \dot{\phi}^2 + {d (d-1)\over4} \dot{\mu}^2
  + {(9-d)(8-d)\over 4}\dot{\nu}^2+{d (9-d) \over 2}\dot{\mu}
  \dot{\nu}
  \nonumber\\
& \qquad \qquad 
 ={1\over 4}e^{2\phi}(d\, p_{d}+(9-d) p_{9-d})+{1\over 4}V(\phi)+V'(\phi) 
\,,
\label{Basic_eq1_dilaton}
\end{align}
where
\begin{align}
  &p_{d}=-{\partial F\over \partial \mu_i} e^{-d\mu-(9-d)\nu}\,,
  ~~ (\forall i=1,\ldots,d)
\\
  &p_{9-d}=-{\partial F\over \partial \nu_i} e^{-d\mu-(9-d)\nu}\,,
  ~~(\forall i=d+1,\ldots,9) 
  \nonumber
\end{align}
in terms of the free energy $F$.

\subsection{\label{Hagedorn_regime}Hagedorn regime}

In the original BV scenario \cite{Brandenberger_Vafa}, the very early universe is compactified over all nine spatial dimensions with radii $R\sim 1$  filled with string gas in thermal equilibrium around the Hagedorn temperature $T_H$ 
\cite{Hagedorn}. 
In this regime, the usual thermodynamic equivalence between the canonical and microcanonical ensembles breaks down and the latter, more fundamental ensembles, must be used. 
Therefore the microcanonical density of states 
$\Omega(E)$ needs to be derived by analyzing the singularities of the one-loop string partition function in the complex $\beta$ plane, where $\beta=1/T$ is the inverse temperature. 
This analysis was carried out in 
\cite{Deo:1988jj,Bassett:2003ck}. 
Following these discussions, the partition function has poles depending on the radii of the universe, and accordingly the equation of state changes.

For small radii, the leading order expression for the temperature and the pressure which come from the density of states 
$\Omega$ is derived in 
\cite{Bassett:2003ck}. 
For large enough $E$, as a result, the temperature remains close to the Hagedorn temperature $T_H$, and the pressure is vanishingly small, 
\begin{eqnarray}
&&T\sim T_H\,, 
\cr
&&P_{d}\sim 0\,,
\cr
&&P_{~9-d}\sim 0\,.
\end{eqnarray}
Therefore, the string gas in the universe with small radii at the Hagedorn temperature can be treated as a pressureless fluid, as expected by $T$-duality.

As the universe grows with larger radii, the above mentioned poles move on the $\beta$ plane. 
This causes the density of states $\Omega$ to change and yields a different temperature and pressure from those of the former ``small radius" regime \cite{Deo:1991mp}. 
There will exist a critical radius $\bar{R}$. 
When the radius is below the critical radius, the universe is still described by a pressureless state at the Hagedorn temperature, while above the value, the temperature and the equation of state are 
\begin{eqnarray}
   &&{1\over T}={\beta_H E -9\over E}\,,
   \cr
   &&P\propto \Bigl({E\over \beta_H E-9}\Bigr)d R^{ d}(E)\,,
\end{eqnarray}
where $d$ denotes the expanding spatial dimensions.
Note that these quantities depend on the total energy $E$ and that it is necessary to solve the equation of motions for $E$ as a function of time numerically. 
It was shown that the radius rapidly expands like an acceleration expansion while the dilaton continues its monotonic decrease 
\cite{Danos:2004jz}.

The above small and large radius phases are the basic states of string gas near the Hagedorn temperature, and the BV mechanism explains how the particular spatial dimensions enter into the second phase. These discussions are, of course, based on the assumption of the thermal equilibrium of string gas and that is the point of our discussion in this paper.

When the energy density decreases and the temperature falls much below the Hagedorn temperature, the string gas will behave as radiation, with $P={1\over d}E$ yielding the radiation dominated evolution, 
$R \propto t^{2/(d+1)}$ in \cite{Bassett:2003ck,Danos:2004jz}. 
In what follows, we discuss the evolution of spacetime and dilaton and, based on the analysis, we study the thermal equilibrium of string gas at the initial Hagedorn regime, where the string gas behaves as pressureless dust.

\section{Time evolution of universe and dilaton: Model I}
We will take the potential of the dilaton to be of two types, each with a different nature. The first example is a simple exponential potential (Model I), while the second is a double-well potential, discussed in the next section (Model II). 

First, we shall discuss the exponential potential case. 
In order to describe the typical behavior 
of the system, we analyze the simple situation, in which all radii are the 
same, $\mu=\nu$. The more generic situation in which 
small spatial dimensions 
($r=e^{\nu}$) and large spatial dimensions 
($R=e^{\mu}$) are separated will be discussed later in this section.


\subsection{Asymptotic solution with exponential potential
\label{secIII.A}}

As a toy model, we discuss a dilaton field with a runaway potential. (See \cite{Berndsen:2005qq} for early works in other context),
\begin{align}
   V(\phi)=b~e^{2a \phi}\,,
   \label{exponential_potential}
\end{align}
where $a$ and $b$ are constant parameters. For simplicity, we assume $\phi\lsim0$ with $a>0$ hereafter to achieve the weak coupling regime $e^{\phi} < 1$.  
Additionally, we take $|b|=1$ since the basic behavior does not change according to the magnitude of $b$. 
The effective potential in the Einstein frame will allow us to intuitively understand the behavior of the dilaton. 
As discussed in Appendix \ref{effective_potential}, the effective potential of the dilaton in the Einstein frame $W(\phi)$ is described as

\begin{align}
W=-e^{\phi\over 2}V\,,
\label{eq:effective potential}
\end{align}
and hence the case with $b<0$ is the usual exponential potential in the Einstein frame. We will also analyze this case. 
For the following two limits, analytic asymptotic solutions can be found.
Let us begin with discussing these solutions before we present numerical solutions.

\subsubsection{$V\to 0$ limit}
\label{subsubsec:V0 limit}

This case is reduced to the standard analysis without the potential term. 
The equations (\ref{eq:eoms1}) with zero pressure give
\begin{eqnarray}
 & & \ddot{\mu}=(2\dot{\phi}-9\dot{\mu})\dot{\mu}+{1\over 4}V' \,,
\cr
 & & \ddot{\phi}=\dot{\phi}^2-{18}\dot{\mu}^2+{1\over 4}V  + V' \,,
  \label{eq:simple form of EOM mu, phi}
\\
 & & E= e^{9\mu -2\phi}(72\dot{\mu}^2-36\dot{\mu}\dot{\phi}+4\dot{\phi}^2+V )\,,
\nonumber
\end{eqnarray}
and then the $V\to 0$ approximation yields
\begin{eqnarray}
  && \frac{d}{dt} \ln \dot{\mu} = \dot{\psi} \,,
\cr
  && \ddot{\psi}=\frac{9}{2}\dot{\mu}^2+\frac{1}{2}\dot{\psi}^2 \,,
  \label{V=0_psi}
\\
&& \frac{d^2}{ dt^2 } (e^{-\psi}) = {E\over 2} .
\nonumber
\end{eqnarray}
Since the total energy is constant, $E=E_0=\mathrm {const,}$ obtained from Eq. (\ref{conserve_E}), the analytic solutions for $\psi(=2\phi-9\mu)$ and $\mu$ are easily obtained from Eq. (\ref{V=0_psi}) as 
\begin{eqnarray}
&&  e^{-\psi} = {E_0\over 4} (t-t_*)^2 + B(t-t_*) + {B^2-9A^2\over E_0}
\,, 
\label{converge_sol}
\\
&& \mu=\mu_*+{1\over {3}}
  \ln \Bigl|{ (E_0 (t-t_*) +2B-6A)(B+{3}A)
          \over 
              (E_0 (t-t_*) +2B+6A)(B-{3}A)}\Bigr|
\,,
\nonumber
\end{eqnarray}
where $t_*$ is the time when the condition $V(\phi)\simeq 0$ is satisfied. 
The integration constants are given by 
$A = \dot{\mu}_*e^{-\psi_*}$,
$B=-\dot{\psi}_*e^{-\psi_*}$, 
where $\mu_*$ and $\psi_*$ are the field values at the time $t=t_*$.

\begin{figure}[t]
\subfigure[ ]{
\includegraphics[width=5cm,angle=-90]{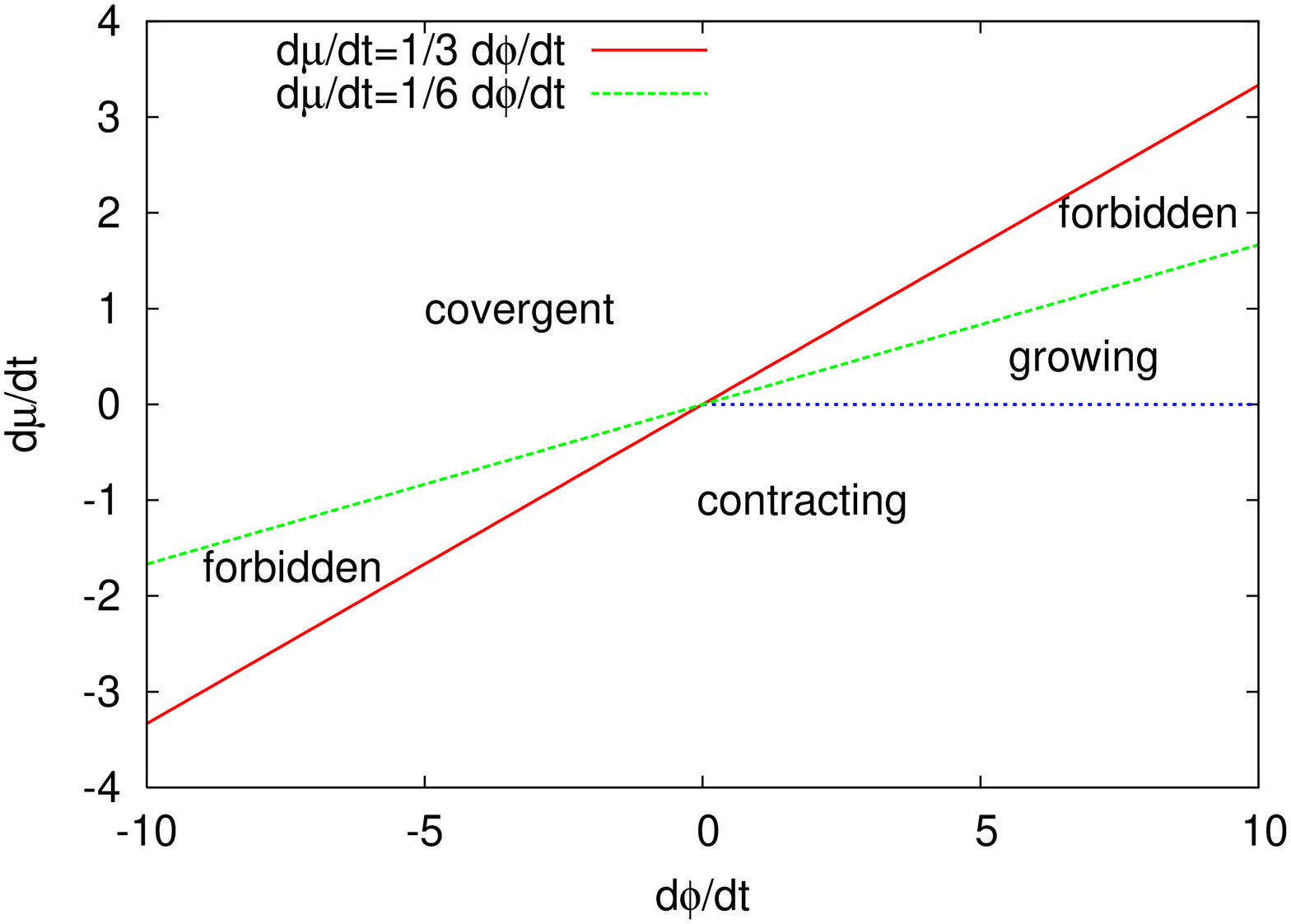}
\label{fig:v=0_sol a}
}
\subfigure[ ]{
\includegraphics[width=5cm,angle=-90]{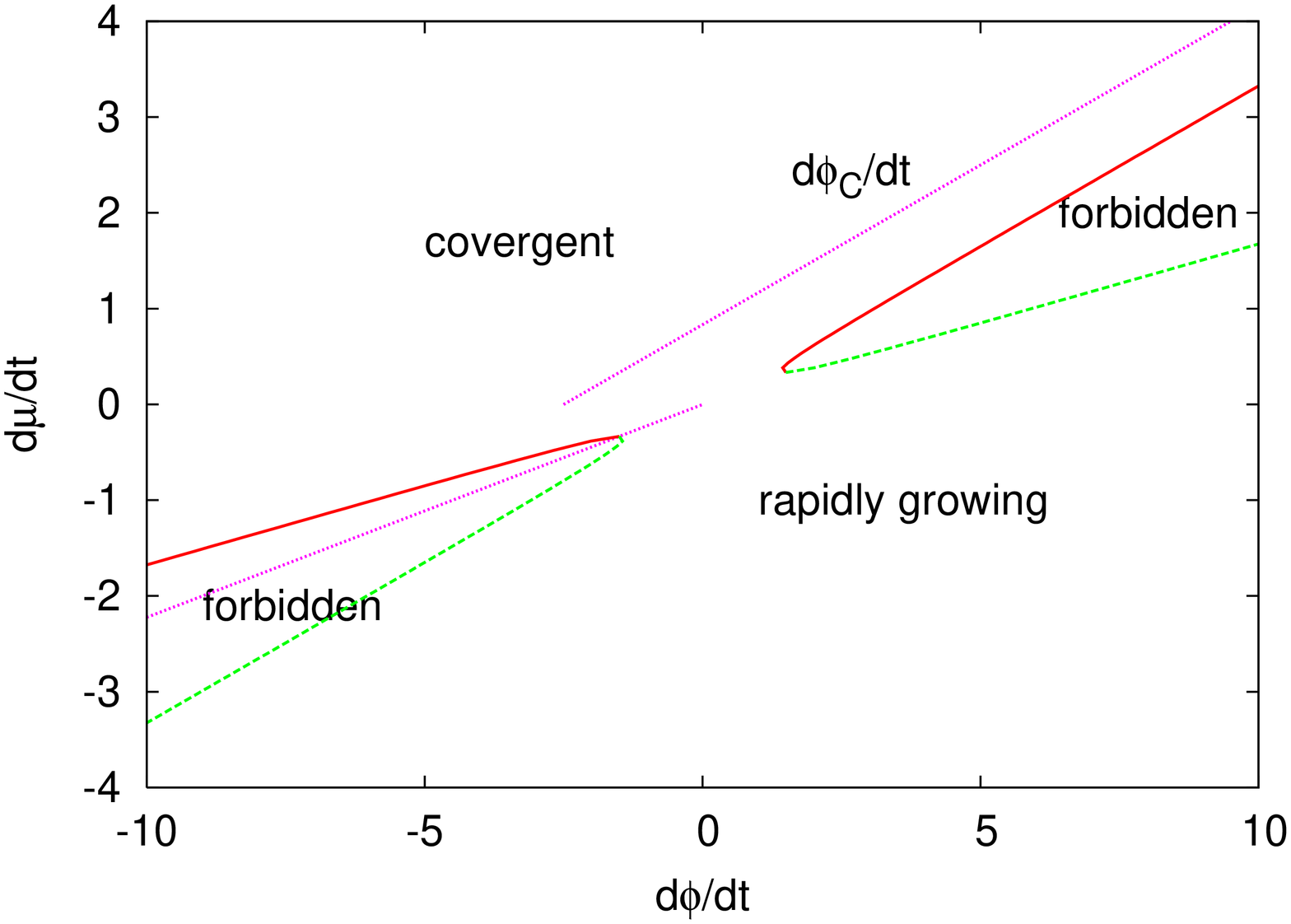}
\label{fig:v=0_sol b}
}
\caption{
Evolution of asymptotic solutions (\ref{converge_sol}) with respect to the initial data $(\dot{\mu}_*, \dot{\phi}_*)$. 
Other initial data are fixed, as $\mu_*=\phi_*=0$ with $t_*=0$.
Fig.~\ref{fig:v=0_sol a} shows the asymptotic solutions with $V=0$. 
The forbidden region $E\lesssim 0$ is equivalent to $3|\dot{\mu}|\lesssim 
|\dot{\phi}|\lesssim 6 |\dot{\mu}|$. 
The growing and contracting solutions correspond to 
$0<\dot{\mu}_*\lesssim \dot{\phi}_*/6$ and $\dot{\mu}_*\lesssim \dot{\phi}_*/3$ with $\dot{\mu}_*<0$, respectively. 
For the other choice of initial data $(\dot{\mu}_*,\dot{\phi}_*)$, the radius converges to a constant value $\mu_{\infty}$ of Eq.~(\ref{asympotic_const}).
Fig.~\ref{fig:v=0_sol b} shows the asymptotic solutions for the exponential potential $V=b e^{2a \phi}$ with $a=2$ and $b=1$. 
The forbidden region $E\lesssim 0$ is equivalent to 
$(9\dot{\mu}-\sqrt{9\dot{\mu}^2-1})/2\lesssim 
\dot{\phi}\lesssim (9\dot{\mu}+\sqrt{9\dot{\mu}^2-1})/2$. 
The rapidly growing and convergent solutions correspond to 
$\dot{\phi}_C<\dot{\phi}_*$ and 
$\dot{\phi}_*<\dot{\phi}_C$, respectively. 
The critical velocity $\dot{\phi}_C$ is approximated by straight lines of Eq.~(\ref{eq:critical velocity}) 
$(\dot{\mu}_*\simeq \dot{\phi}_*/3+5/6)$ with $\dot{\mu}_*>0$ and 
Eq.~(\ref{eq:critical velocity H<0}) $(\dot{\mu}_*\simeq 2/9\dot{\phi}_*)$ with $\dot{\mu}_*<0$ 
(Appendix B). }
\label{fig:v=0_sol}
\end{figure}

The solutions behave asymptotically in three ways (see Fig.~\ref{fig:v=0_sol a}).  
The first type (i) is that in which the radius grows as $\lim_{t\to 0} \ln |1/t|$ and diverges ($\mu\to \infty$) at $E_0(t-t_*)=2|B+3A|$. At the same time, the dilaton also grows and diverges 
\footnote{
At the strong-coupling regime after the growth of the dilaton, we cannot predict what will happen. A possibility of a static universe in the strong-coupling regime is recently discussed in \cite{Brandenberger:2006pr}.
}.
 These behaviors appear 
 if the conditions $B+3A<0$ and $|B+3A|<|B-3A|$ are satisfied, yielding 
 $0<6\dot{\mu}_*< \dot{\phi}_*$. 
The second type of evolution (ii) is that in which the radius contracts as 
 $\lim_{t\to 0} \ln |t|$ and $|\mu|$ diverges at $E_0(t-t_*)=2|B-3A|$. 
Similarly, the dilaton contracts. These behaviors appear if the conditions $B-3A<0$ and $|B-3A|<|B+3A|$ are satisfied, yielding 
 $3\dot{\mu}_*< \dot{\phi}_*$ with $\dot{\mu}_*<0$. 
The last type (iii) is that in which the radius converges to a constant value, which is achieved for $0<B+3A, B-3A$.

The first two types of evolution will be seen in the double-well potential case (Sec.~\ref{sec:Model II}). 
In the present model, only the last behaviors are important and we comment on its asymptotic radius. The asymptotic evolution at $t\to \infty$ of the spacetime is 
\begin{align}
  R_{\infty} =e^{\mu_{\infty}} = e^{\mu_*}\Bigl|{B+{3}A\over B-{3}A}
              \Bigr|^{1/{3}}\,,
              \label{asympotic_const}
\end{align}
and it converges to a constant radius \cite{Tseytlin:1991xk}, while the (shifted) dilaton rolls monotonically to the weak coupling.
Without fine-tuning ($B\simeq 3A$), the asymptotic value 
$R_{\infty}$ is not very large, and the radius remains small. 
We call this solution {\it{the convergent solution}}.

\subsubsection{$V,V'\gg \dot{\phi}\dot{\mu}, ~ \dot{\phi}^2,~ \dot{\mu}^2$ limit}

In the potential dominated case, the equations 
(\ref{eq:simple form of EOM mu, phi}) are reduced to 
\begin{eqnarray}
   &&\ddot{\mu}=\frac{1}{4}V'(\phi)= b \frac{a}{2}e^{2a\phi}
   \,,
   \cr
   &&
   \ddot{\phi}=\frac{1}{4}V(\phi)+V'(\phi) = b
   \frac{8a+1}{4}e^{2a \phi}\,.
   \label{case_b_2}
\end{eqnarray}
As for $b=1$, if the initial condition satisfies 
\begin{eqnarray}
    {8a+1\over 4a}e^{2a\phi_*} >  \dot{\phi}_*^2 \,,
\label{dot_phi_small}
\end{eqnarray}
the analytic solutions for these equations are given by

\begin{eqnarray}
 && \phi=-\frac{1}{a}\ln \Biggl\{  \frac{1}{2C_1}  \sqrt{ 8 + \frac{1}{a}}
  \sin \Bigl[a C_1 \bigl(  t-t_* +C_2 \bigr) \Bigr]\Biggr\}
  \,,
  \cr
  && \mu={2a\over 8a+1}\phi+D_1 (t-t_*) +D_2 .
  \label{diverge_sol}
\end{eqnarray}
Here $C_1,C_2,D_1$ and $D_2$ are the integration constants given by
\begin{eqnarray} 
&& C_1 = \sqrt{{8a+1\over 4a}e^{2a\phi_*}-\dot{\phi}_*^2} \,,
\cr
&& C_2 ={ 1 \over a C_1 } \cot^{-1} \biggr| \frac{\dot{\phi}_*}{C_1} \biggr|\,,
\cr
&& D_1 =\dot{\mu}_*-{2a\over 8a+1}\dot{\phi}_*\,,
\cr
&& D_2 =\mu_*-{2a\over 8a+1}\phi_*\,.
\label{eq:D1 D2}
\end{eqnarray}
$\mu_*$ and $\phi_*$ are the field values at the time $t=t_*$ 
when  the condition 
$V,V' \gg \dot{\phi}\dot{\mu}, \dot{\phi}^2, \dot{\mu}^2$ becomes a good approximation. 
Both the dilaton and radius diverge at
\begin{align}
   ( t-t_* ) \sim {\pi \over a C_1} - C_2 \,.
   \label{eq:divergent point}
\end{align}
We call the solution (\ref{diverge_sol}) {\it{the rapidly growing solution}}. 
Note that for  $ {8a+1\over 4a}e^{2a\phi_*} < \dot{\phi}_*^2$ the solutions are given by Eqs.~(\ref{diverge_sol_ver2}) with  $b=1$. 
\footnote{
However, this solution does not appear in the numerical simulation in the next subsection because the dilaton $\phi$ goes to $-C_1 t$, violating the 
$V$-dominated condition. This case is, as a result, reduced to $V\to0$ limit in Sec.~\ref{subsubsec:V0 limit}.}

For $b=-1$, the solutions become
\begin{eqnarray}
&& \phi=-\frac{1}{a}\ln \Biggl\{  \frac{1}{2C_1}  \sqrt{ 8 + \frac{1}{a}}
  \sinh \Bigl[a C_1 \bigl(  t-t_* +C_2 \bigr) \Bigr]\Biggr\}
  \,,
  \cr
&& C_1 = \sqrt{  \dot{\phi}_*^2 - b {8a+1\over 4a}e^{2a\phi_*} } \,.
\label{diverge_sol_ver2}
\end{eqnarray}
The dilaton asymptotes to $\phi \to - C_1 t$, toward the weak coupling, and at the same time the radii inflate if 
$(8a+1) \dot{\mu}_* /[2a (\dot{\phi}_* + C_1)] \gtrsim 1 $; otherwise the radii contract exponentially. 
The contracting evolution appears in the numerical simulation since the condition for the rapidly contracting  
$(8a+1) \dot{\mu}_* /[2a (\dot{\phi}_* + C_1)] \lsim 1 $ can be initially satisfied for $|\dot{\phi}_0|\sim O(1)$ and $\dot{\mu}_0\lsim O(1)$.

\subsection{Numerical analysis}
\label{sec:numerical analysis model I}

\begin{table}[tb]
\begin{ruledtabular}
\begin{tabular}{c|lllllc}
Model & Parameter & & Type of solution \\
\hline
& $b>0$ & $a>1$
& convergent solution ($\dot{\phi}_0 < \dot{\phi}_C$)
\\
&   &
& rapidly growing solution ($\dot{\phi}_0 > \dot{\phi}_C$)
\\
$V=b e^{2a \phi}$
&    & $a \le 1$ 
& rapidly growing solution
\\
&  $b<0$ & $a >1$
&  convergent solution ($\dot{\phi}_0 < \dot{\phi}_C$)
\\
&         &
&  contracting solution ($\dot{\phi}_0 > \dot{\phi}_C$)
\\
&         &
&  convergent solution ($H_0>0$ and $\dot{\phi}_0 \lesssim 0$)
\\
\hline
\\
$V= \frac{\lambda}{4}\Bigl[ \left( \frac{\phi}{\eta}\right)^2-1\Bigr]^2$ 
&  $\lambda>0$ & $\eta \gg 1$
&  rapidly growing solution 
   ($\dot{\phi}_0 \lesssim 0$)
\\
&  $\lambda<0$ &
&  stabilized dilaton with $\phi \to -\eta$.
   ($H_0>0$ and $\dot{\phi}_0 \lesssim 0$)
\\
\end{tabular}
\end{ruledtabular}
\caption{
Summary of basic behaviors
\label{table:flowchart}
}
\end{table}

Now we are ready to present numerical solutions of the basic equations 
(\ref{eq:simple form of EOM mu, phi}).
We first discuss the case in which all spatial dimensions have the same radii, 
$\lambda_i=\mu$. 
The other case with $\lambda_i \neq \mu$ will be discussed in the next subsection.

In order to see the typical evolutions of dilaton and spacetime, we change the initial velocity $\dot{\phi}_0$ for various values of $a$, with respect to the fixed  initial data of $\mu_0$, $\phi_0$ and $\dot{\mu}_0$ (Appendix \ref{phase_plane}). 
We are interested in the situation where all initial radii are near the string scale, $R \sim 1$, and where they expand (or contract) with time. 
Thus, it is natural to consider $\mu_0 \approx 0$ and 
$ |\dot{\mu}_0| < 1$, where the latter condition comes from the adiabaticity of the cosmological expansion. In our discussion, let us follow Table 
\ref{table:flowchart}.

\subsubsection{$b>0$ case}

\begin{center}
\it{Initially expanding case ($H_0>0$)} 
\end{center}

We begin with studying the initially expanding case, 
$H_0=\dot{\mu}_0>0$. 
For the initial condition of the dilaton, we consider 
$\phi_0 \lsim0$ and 
$\dot{\phi}_0<0$, which satisfy the weak coupling condition. 
The case of $\dot{\phi}_0>0$ will be discussed soon hereafter.

\paragraph{Initial data of $\dot{\phi}_0<0$.}

\begin{figure}[t]
\begin{center}
\scalebox{0.36}{\includegraphics[angle=-90]{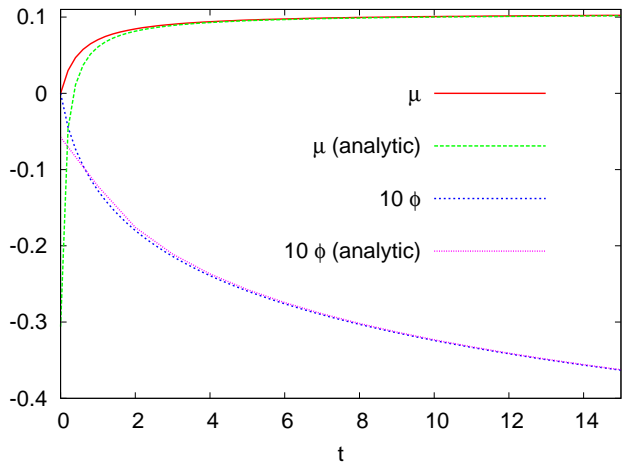}}
\caption{
Plot of typical convergent solutions of the dilaton $\phi$ (blue line) and radius $R=e^{\mu}$ (red line) for $\mu_0=\phi_0=0, 
\dot{\mu}_0=0.2,\dot{\phi}_0=-3$ and $a=2,b=1$. 
The analytic solutions (green and pink lines) of Eq. (\ref{converge_sol}) are also plotted as a reference.
Here the integration constants, $\mu_*,\phi_*,\dot{\mu}_*,\dot{\phi}_*$, are chosen appropriately. 
The asymptotic evolutions are well approximated by the analytic solutions. 
The same results can be obtained for the case $\dot{\phi}_0>0$ with 
$\dot{\mu}_0 \gtrsim O(1)$.}
\label{converge_com}
\end{center}
\end{figure}

\begin{figure}
\begin{center}
\scalebox{0.36}{\includegraphics[angle=-90]{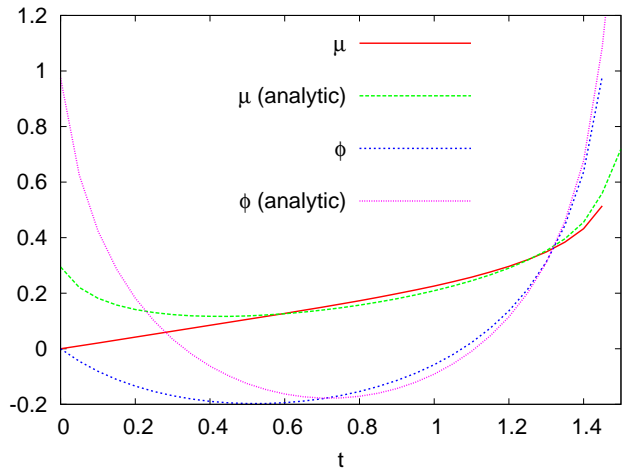}}
\caption{Plot of typical rapidly growing solutions 
of the dilaton and radius  for the same initial conditions in 
Fig. \ref{converge_com} except for $\dot{\phi}_0=-1$.}
\label{diverge_com}
\end{center}
\end{figure}

We set these initial conditions for various values of $a$ and numerically solve Eqs. 
(\ref{eq:simple form of EOM mu, phi}). 
For $a>1$ we find that there is a 
critical value of $\dot{\phi}_0$, denoted $\dot{\phi}_C$. 
The critical velocity $\dot{\phi}_C$ is approximated by a 
straight line in the $(\dot{\mu}_0,\dot{\phi}_0)$ plane, 
and the approximated line will be 
obtained by Eq. (\ref{eq:critical velocity}) in Appendix 
\ref{phase_plane} (Fig. \ref{fig:v=0_sol b}), 
where we will find 
$\dot{\phi}_C<0$ for $\dot{\mu}_0\lsim O(1)$. 
For $\dot{\phi}_0<\dot{\phi}_C$, 
the radius goes asymptotically to a constant value, 
and the dilaton monotonically decreases, 
while both radius and dilaton diverge 
at late times for $\dot{\phi}_0>\dot{\phi}_C$. 
These two typical evolutions may be naturally 
related to the analytic solutions in the previous subsection. 
We plot these results in Figs. \ref{converge_com} and 
\ref{diverge_com}, comparing them with the analytic solutions.
Fig. \ref{converge_com} shows a typical result for
$\dot{\phi}_0\lsim \dot{\phi}_C$, 
which corresponds to the convergent solution (\ref{converge_sol}), and 
Fig. \ref{diverge_com} shows 
the opposite case, which is related to 
the divergent solution (\ref{diverge_sol}), 
with appropriate choice of integration constants, 
$\{ \mu_*,\phi_*,\dot{\mu}_*,\dot{\phi}_* \}$.

For any value of $\dot{\phi}_0 < \dot{\phi}_C$, 
the behaviors of numerical solutions are the same as in Fig. 
\ref{converge_com}. Similarly, for any value of 
$\dot{\phi}_0 > \dot{\phi}_C$ the behaviors are the same as Fig. 
\ref{diverge_com}. 
For $\dot{\phi}_0 \sim \dot{\phi}_C$, the radius seems to approach a constant value at first, but $\ddot{\mu}$ changes its sign from minus to positive at late times, and then $\mu$ eventually diverges. 
All these asymptotic evolutions are well approximated 
by the analytic solutions
\footnote{
It is also interesting to see if the numerical solutions satisfy the condition for the rapidly growing solution 
Eq. (\ref{dot_phi_small}) at the initial time $(t=0)$.
The condition Eq. (\ref{dot_phi_small}) 
is initially satisfied for $\dot{\mu}_0\gsim 0.3$ 
with $a=2$, while for $\dot{\mu}_0\lsim 0.2$ 
the numerical solutions initially 
violate the condition Eq. (\ref{dot_phi_small}), but 
they immediately become the rapidly growing ones.}.

For $0<a\le 1$, 
the critical velocity and the convergent solution disappear, 
and all numerical solutions become the rapidly growing solutions, as in 
Fig. \ref{diverge_com}.


\paragraph{Initial data of $\dot{\phi}_0>0$.}

At first sight, solutions with $\dot{\phi}_0>0$ seem to easily violate the weak coupling condition. 
However, there are particular sets of parameters 
$(\dot{\mu}_0,\dot{\phi}_0)$ that preserve the weak coupling condition. 
The same condition described above is also applicable in this case: for any value of $\dot{\phi}_0 <\dot{\phi}_C$, the numerical solutions are asymptotically the convergent solutions, as in Fig. \ref{converge_com}, while for $\dot{\phi}_0 >\dot{\phi}_C$,  the numerical solutions are asymptotically the rapidly growing ones as in 
Fig. \ref{diverge_com}. 
For $\dot{\phi}_0>0$, the convergent solution is allowed; otherwise, the weak coupling condition is immediately violated. This is realized by a positive critical velocity larger than $\dot{\phi}_0$ ($\dot{\phi}_C>\dot{\phi}_0>0$), yielding large initial velocity of radius $\dot{\mu}_0\gsim 1$ 
(Fig. \ref{fig:v=0_sol}). 
However, this case violates the adiabatic condition $\dot{\mu}\gsim O(1)$.

\begin{center}
\it{Initially contracting case ($H_0<0$)}
\end{center}

Next, we discuss the initially contracting case, $H_0=\dot{\mu}_0<0$. 
In this case, we also find that the asymptotic behavior is the same as 
$H_0>0$, even though the radius initially contracts. 
The critical velocity $\dot{\phi}_C$ is also approximated by a straight line (see Eq. (\ref{eq:critical velocity H<0})).  
The slope is different from that of 
$H_0>0$, and we find $\dot{\phi}_C$ is negative in the present case 
(Fig. \ref{fig:v=0_sol b}). 
Hence for $\dot{\phi}_0<0$, any initial velocity of $\dot{\phi}_0 <\dot{\phi}_C$ yields the convergent solution, and the opposite case yields the rapidly growing one. On the other hand, for $\dot{\phi}_0>0$, all numerical solutions become the rapidly growing ones since $\dot{\phi}_C<0$.

In summary, we find that for $a>1$, there are mainly two asymptotic behaviors, i.e., either convergent or rapidly growing, depending on the initial dilaton velocity $\dot{\phi}_0$ with respect to the other fixed initial values. 
On the other hand, for $a\leq 1$, we do not find evolution like the convergent solution. 
This feature will be understood clearly in a phase-space analysis in Appendix \ref{phase_plane}.

\subsubsection{$b<0$ case}
\begin{center}
\it{Initially expanding case ($H_0>0$) }
\end{center}

\paragraph{Initial data of $\dot{\phi}_0<0$}
First, we consider the initially expanding case with $\phi_0\lsim 0$ and
$\dot{\phi}_0<0$. 
For $a>1$, we find that there is a critical value 
$\dot{\phi}_C$, similar to the case where $b>0$. 
The asymptotic behaviors are, however, different from the previous $b>0$ case.  
For $\dot{\phi}_0<\dot{\phi}_C$, the radius asymptotes to a constant value and the dilaton monotonically decreases, while both radius and dilaton contract exponentially for 
$\dot{\phi}_0>\dot{\phi}_C$
\footnote{
The exponentially contracting phase is not interesting from the point of view of decompactification; thus 
we will mainly pay attention to the convergent solution. }.

These two typical evolutions are related to the analytic solutions of Eqs. (\ref{converge_sol}) and (\ref{diverge_sol_ver2}) in 
Sec. \ref{secIII.A}. 
Because 
$\dot{\phi}_C$ is positive in the initially expanding case, 
the solutions become the convergent for all initial conditions 
$\dot{\phi}_0<0$.

Notice that the forbidden region, $E\lsim 0$ for 
the $b=-1$ case, which is equivalent to the condition 
$ (9\dot{\mu}_0-\sqrt{1+9\dot{\mu}_0^2})/2\lsim \dot{\phi}_0\lsim 
(9\dot{\mu}_0+\sqrt{1+9\dot{\mu}_0^2})/2$, exists near the critical velocity. 
So the critical velocity $\dot{\phi}_C$ is the same as either above boundary value of this forbidden region. 
The critical velocity $\dot{\phi}_C$ is also approximated by a straight line in the $(\dot{\mu}_0,\dot{\phi}_0)$ plane as in the case where $b>0$. The slope is different from $b>0$ and the above forbidden band $E\lsim0$ is near the line.

\paragraph{Initial data of $\dot{\phi}_0>0$}

The same condition as above is also applicable in this case: for any value of $\dot{\phi}_0<\dot{\phi}_C$, the behaviors of numerical solutions are asymptotically the convergent solutions, while for any value of $\dot{\phi}_0>\dot{\phi}_C$, the numerical solutions are asymptotically the rapidly contracting ones. 
Although  $\dot{\phi}_0>0$ seems to violate the weak coupling, the weak coupling condition is held asymptotically since the dilaton goes to the weak coupling region for both the convergent solution and the contracting one.

\begin{center}
\it{Initially contracting case ($H_0<0$)}
\end{center}

Next we discuss the initially contracting universe. 
In this case, we also find that the asymptotic behavior is the same as $H_0>0$, even though the radius initially contracts. 
The critical velocity $\dot{\phi}_C$ in the present case is also approximated by a straight line, and we find $\dot{\phi}_C$ is negative.
Hence for $\dot{\phi}_0<0$, any lower value 
$\dot{\phi}_0<\dot{\phi}_C$ yields the convergent solution and the opposite case yields the rapidly contracting one. 
On the other hand, for $\dot{\phi}_0>0$, all numerical solutions become the rapidly contracting ones because $\dot{\phi}_C<0$. 
Notice that the critical velocity $\dot{\phi}_C$ is also the same as either boundary value of the forbidden region of  $E\lsim 0$, 
similar to the case where $H_0>0$.

In summary, we find that for $a>1$ (and $b<0$), there 
are mainly two asymptotic behaviors, i.e., 
the convergent or rapidly contracting, depending on the 
initial dilaton velocity $\dot{\phi}_0$ with 
respect to the other fixed initial values
\footnote{
The classification of the solution for $a \le 1$ is more complicated.
For the initially expanding case with $1/2 < a\le 1$, 
in addition to the above two evolutions, 
there is the rapidly growing solution for the very large velocity 
$\dot{\phi}_0\gg 1$. 
As for $a\le 1/2$ the contracting solution 
disappears. Hence the asymptotic solutions are 
of two types: the convergent and rapidly expanding ones. 
However, the rapidly expanding solution in the present case initially violates the weak coupling condition because $\dot{\phi}_0>0$. 
On the other hand, for the initially contracting case with $a\le 1$, 
there exists convergent and contracting solutions similar to those where 
$a> 1$. This $a$-dependence will be clarified in Appendix \ref{phase_plane} 
by employing phase-space analysis. 
}.
In the initially expanding case, for $\dot{\phi}_0\lsim0$, all numerical solutions become convergent solutions.  
The behavior will be easily understood in the effective potential picture in the next subsection.

\subsubsection{Effective potential picture}

As mentioned above, the effective potential of the dilaton in the Einstein frame is $W=-b\,e^{2\til{a}\phi}$ with a redefined constant $\til{a}=a+1/4$ obtained from Eq. (\ref{eq:effective potential}). 
Adopting the Einstein frame, we can explain intuitively why the evolutions depend on an initial velocity $\dot{\phi}_0$.

\begin{center}
\it{$b>0$ case}
\end{center}

The effective potential has a valley in the strong coupling regime ($\phi>0$) and asymptotes to a flat value, $W\to 0$, in the weak coupling regime ($\phi<0$). 
In this picture, when the dilaton starts from the origin $\phi_0\lsim 0$, with a small initial velocity, it rapidly falls down a positive side of the potential. 
On the other hand, if it starts with a sufficiently large velocity to avoid falling, it can climb up the valley of potential toward $W=0$. 
Therefore, $V$ becomes dominant for small initial velocities, while for large initial velocities, $V$ goes to zero. 
These behaviors are consistent with the above-mentioned two regimes in the analytic method.

\begin{center}
\it{$b<0$ case}
\end{center}

This case is equivalent to the usual picture of exponential potential in terms of $W$
\footnote{As a natural potential for the dilaton, 
$W\propto e^{e^{\phi}}$  
is interesting to study \cite{Brustein:1992nk,Dine:2000ds}.  
It physically corresponds to the case $a\gg 1$ with $b=-1$ in our setup. 
However, the basic behaviors do not 
change and are similar to the case where $a>1$ described above. 
In fact, we have confirmed that the basic behaviors of the dilaton 
with $W=e^{e^{\phi}}$ are similar to the case where $a\gg 1$.},
where the dilaton rolls toward $W=0$ in the weak coupling.
In this picture, when the dilaton starts from the origin $\phi_0\lsim 0$, 
it tends to go to $V=0$, asymptoting to $\phi\to -\infty$. 
For the initial condition,  
$\dot{\phi}_0<0$ or $0<\dot{\phi}_0\lsim O(1)$, the solutions become $V\to 0$, while for $\dot{\phi}> O(1)$,  
$\mu$ begins to rapidly contract since the condition $(8a+1) \dot{\mu}_* /[2a (\dot{\phi}_* + C_1)] \lsim 1 $ is satisfied  before $V$ goes to zero, 
as shown in the previous discussion.

\subsection{Evolution of large and small radii}

So far, we have assumed all spatial directions have the same radius.
As expected, the basic behaviors observed under such simplification hold even when there is asymmetry between several spatial radii.
We briefly discuss such cases, representing a typical example of evolution with different spatial radii.

\begin{figure}[t]
\begin{center}
\scalebox{0.36}{\includegraphics[angle=-90]{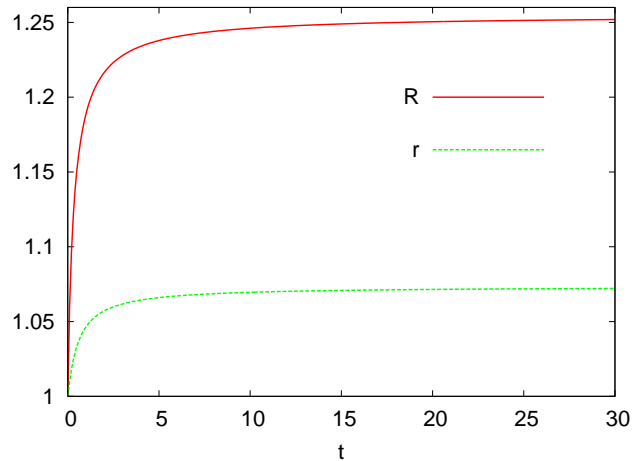}}
\caption{
Typical convergent solutions of radii of large spatial $d=3$ dimensions 
($R=e^{\mu}$) and small 
$6$ dimensions ($r=e^{\nu}$).
The initial conditions are $\mu_0=\nu_0=\phi_0=0, \dot{\mu}_0=0.8, \dot{\nu}_0=0.1,\dot{\phi}_0=-3$ and 
$a=b=1$.}
\label{each_dim1_ex}
\end{center}  
\end{figure}
\begin{figure}[t]
\begin{center}
\scalebox{0.36}{\includegraphics[angle=-90]{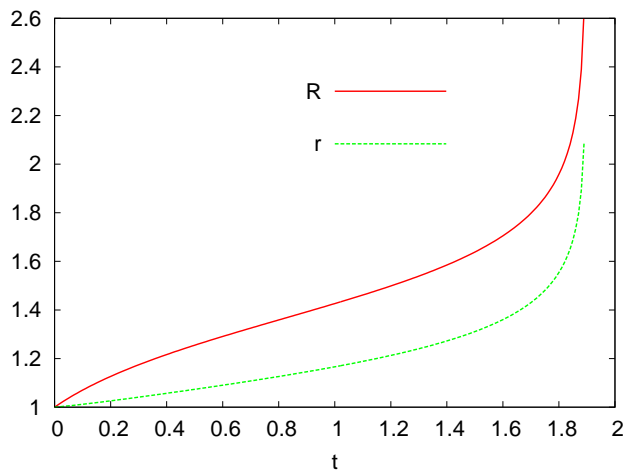}}
\caption{
Typical rapidly growing solutions of 
$R=e^{\mu}$ and $r=e^{\nu}$ for the same initial conditions and 
parameters as in Fig. \ref{each_dim1_ex} except for $\dot{\phi}_0=-1$.
}
\label{each_dim2_ex}
\end{center}  
\end{figure}

\begin{center}
\it{$b>0$ case}
\end{center}

In Figs. \ref{each_dim1_ex} and \ref{each_dim2_ex}, we plot typical two types of time evolution of radii, as a solution of (\ref{Basic_eq1_dilaton}) for $d=3$ with the potential (\ref{exponential_potential}). 
From these figures, it is understood that each radius evolves in the same way: both of them are either convergent solutions or rapidly growing ones. 
We could not find numerical solutions in which the radii show different types of behavior in late times. 
The reason why such solutions are not found is as follows. 
Suppose the potential $V$ dominates in the equation of motion for 
$\mu$, then $\nu$ becomes the convergent solution asymptotically. 
This assumption makes $\mu$ rapidly grow. 
In this case, the derivatives of $\nu$ are asymptotically vanishing, and by substituting this condition into the second of 
Eqs. (\ref{Basic_eq1_dilaton}), we obtain the equation $V'\simeq 0$. 
This is inconsistent with our assumption that $V$ is dominant. 
Therefore the rapidly growing solution of  $\mu$ is inconsistent with the convergent solution of $\nu$, and vice versa.

We conclude that for $a>1$ the behavior of each radius is determined by the initial velocity $\dot{\phi}_0$ once $\dot{\mu}_0$ and $\dot{\nu}_0$ are fixed.  
Both radii evolve as the rapidly growing solutions for a large dilaton velocity while they are the convergent solutions for a small velocity.

\begin{center}
\it{$b<0$ case}
\end{center}

Both $\mu$ and $\nu$ behave in the same way as they do in the case where $b>0$: both radii are convergent solutions or rapidly contracting ones since the same argument for $b>0$ is applied in the present case, except for the different evolution during the $V$-dominated regime. 
For $a>1$, $\dot{\phi}_0<0$ and $0<\dot{\mu}_0, \dot{\nu}_0\lsim O(1)$, both radii always become convergent solutions, and the typical evolutions are the same as in Fig. \ref{each_dim1_ex}.

\section{Time evolution of universe and dilaton: Model II}
\label{sec:Model II}

\subsection{\bf Double-well potential and asymptotic solutions}

As a second model, we consider the double-well potential, 
\begin{align}
     V(\phi)={\lambda\over 4}\Biggl[\Bigl({\phi\over \eta}\Bigr)^2-1\Biggr]^2\,,
\end{align}
where $\lambda$ and $\eta$ are a coupling constant and a vacuum expectation value (VEV) of $\phi$, respectively. 
Similar to the previous model, $\lambda$ is taken to be positive or negative.
 The later case corresponds to an ordinary concave effective potential picture $W\sim -V$ (Appendix \ref{effective_potential}). 
For simplicity, we assume until \S \ref{sec: IV C} that all radii are the same, $\mu=\nu$.

For $\lambda<0$
\footnote{For the case where $\lambda>0$, we find solutions 
like Eqs. (\ref{stab_in d-w}), which represent the dilaton oscillating around $\phi=0$. 
},
the dilaton stays at the minimum (VEV) of the double-well potential 
$\phi_V=\pm \eta$. 
Expanding $\phi$ around the VEV,  
\begin{eqnarray}
&&V \approx O (\delta \phi^2)\,,
\cr
&&V' \approx  {2\lambda \over \eta^2}\delta \phi + O (\delta \phi^2)\,,
\end{eqnarray} 
we find that Eqs. (\ref{eq:simple form of EOM mu, phi}) are approximated as 
\begin{eqnarray}
&&\ddot{\mu}+9\dot{\mu}^2\approx 0\,,\cr
&&\delta \ddot{\phi}
 - \frac{2 \lambda}{\eta^2}  \delta {\phi}
 + 9 \dot{\mu} \delta \dot{\phi}+{E\over 4} e^{-9\mu + 2\phi_{V}} 
 \approx 0\,.
\label{eq:approx sol modelII}
\end{eqnarray}
These equations have the following solutions: 
\begin{eqnarray} 
 &&\delta \phi \sim 
  C_1 J_{0}(z)+C_2 N_{0}(z)\,,
\cr
 &&\mu\sim D_2+{1\over 9}\ln\Bigl|t-t_*-{D_1\over 9}\Bigr|
 \,, \label{stab_in d-w}
\\
  && z={\sqrt{2|\lambda|}\over \eta}\Bigl(t-t_*-{D_1\over 9}
  \Bigr)\,, 
\nonumber
\end{eqnarray}
where $J_{\nu}(z),N_{\nu}(z)$ are the Bessel functions with the amplitude 
$C_1$ and $C_2$. $D_1$ and $D_2$ are given by 
\begin{eqnarray}
&&D_1=-{1\over \dot{\mu}_*}\,,
\nonumber
\\
&&D_2=\mu_*+{1\over 9}\ln 9 |\dot{\mu}_*|\,.
\end{eqnarray}
Note that when we derive the above solutions, we have used the fact that the last term in (\ref{eq:approx sol modelII}) decays as $e^{-9\mu + 2\phi_{V}} \propto 1/t$, so that it can be omitted at late times. 
The amplitude of $\delta \phi$ becomes 
$\propto 1/\sqrt{t}$ for $\lambda<0$, which is equivalent to a decaying solution. 
We will call this solution {\it the stabilized dilaton solution.}

\subsection{\bf Numerical analysis} 

Now we are ready to present the numerical solutions of the system. We take the initial condition in such a way that 
$\mu_0(\sim 0)$, $\phi_0(\lsim 0)$ and $\dot{\mu}_0$ are fixed and $\dot{\phi}_0$ is changed for various values of $\lambda$ and $\eta$, within the adiabatic condition. 
The evolutions depend on the sign of $\lambda$, and so we begin with the case $\lambda>0$, following Table~\ref{table:flowchart}.

\subsubsection{$\lambda>0$ case}

We find that the potential is negligible and hence does not work for the large initial velocities $\dot{\phi}_0>3\dot{\mu}_0$ and $\dot{\phi}_0>6\dot{\mu}_0$ with the initially expanding and contracting case, which are equivalent to the regions labeled as ``growing"  and ``contracting" in Fig.~\ref{fig:v=0_sol a}. 
For these initial conditions, the numerical solutions are the same as the analytic solutions (\ref{converge_sol}) with 
$t_*=t_0$, obtained by $V\to 0$ approximation. As discussed in \S III. A, the asymptotic behaviors are two types: both the radius and dilaton either grow or contract. 
However, from the point of view of decompactification, the contracting solutions are not interesting. 
Moreover, the growing solutions violate the weak coupling condition under $\phi_0\sim 0$ due to the positive velocity, $\dot{\phi}_0(>6\dot{\mu}_0)>0$. 
Therefore, we discuss the other initial conditions, 
$\dot{\phi}_0<3\dot{\mu}_0$ (or $\dot{\phi}_0 < 6\dot{\mu}_0$), for which the potential works.

In this case, there exists a critical value of 
$\dot{\phi}_0$, denoted $\dot{\phi}_C$, which determines the evolutions as follows. The radius begins to contract at late times as the dilaton rolls to the negative infinity for 
$\dot{\phi}_0<\dot{\phi}_C$, while the radius diverges at late times as the dilaton rolls to the positive infinity for 
$\dot{\phi}_0>\dot{\phi}_C$. 
This behavior is easily understood by employing effective potential picture.
The effective potential of the dilaton is given by 
$W\sim -V$. In this picture, for $\lambda>0$, there exist two valleys in both sides of the potential for $|\phi|\to \infty$. 
When the dilaton starts from the origin, it asymptotically goes to the positive (negative) infinity, 
$|\phi| \to \infty$, if the initial velocity is  $\dot{\phi}_0>\dot{\phi}_C$ ($\dot{\phi}_0 < \dot{\phi}_C$). 
On the other hand, taking $\phi,V\gg 1$ limit yields
$V\approx \lambda/4(\phi/\eta)^4$. 
Combining this with the fact that the total energy is constant from (\ref{conserve_E}), 
$E\approx e^{9\mu-2\phi}V\simeq {\rm const}$, we obtain $e^{9\mu}\propto e^{2\phi}\phi^{-4}$. 
This equation means that the dilaton going to the positive (negative) infinity makes the radius go to the positive (negative) infinity ($|\mu|\to \infty$). 
$\mu \to \infty$ corresponds to the rapidly growing solution, and $\mu \to - \infty$ corresponds to the contracting solution.

Even though the critical velocity $\dot{\phi}_C$ depends on both parameters $\lambda$ and $\eta$, it depends more sensitively on $\eta$ than $\lambda$.  
The value of $\dot{\phi}_C$ becomes smaller as $\lambda$ or $\eta$ grows. 
At $\eta\sim 1$, $\dot{\phi}_C$ is positive $\sim O(1)$. 
Then for $\eta\lsim 1$ with $\dot{\phi}_0<0$, in which the weak condition will be satisfied, the radius always begins to contract since the condition $\dot{\phi}_0<\dot{\phi}_C$ is satisfied. 
Since we are mainly interested in the expanding phase, let us consider $\eta> 1$. 
For a large value of $\eta(\gg 1)$, e.g., $\eta\gsim5$, both the radius and dilaton expand and diverge at some late times for any value of $\dot{\phi}_0<0$ since $\dot{\phi}_C( <-O(30))$ is sufficiently small to satisfy the condition for the rapidly growing solutions, $\dot{\phi}_0>\dot{\phi}_C$. 
It means the dilaton goes to the positive infinity for a very wide range of parameters. We will revisit this behavior soon later (Sec. \ref{subsubsec:effective picture modelII}).

In summary, we find that there are mainly two asymptotic behaviors, i.e., contracting or rapidly growing, depending on the values of initial dilaton velocity $\dot{\phi}_0$.
In the case $\dot{\phi}_0<0$, in which the weak coupling is satisfied, all numerical solutions become the rapidly growing ones for both the initially expanding and contracting cases as long as $\eta \gg 1$.

\subsubsection{$\lambda<0$ case}

\begin{figure}[htmp]
\begin{center}
\scalebox{0.36}{\includegraphics[angle=-90]{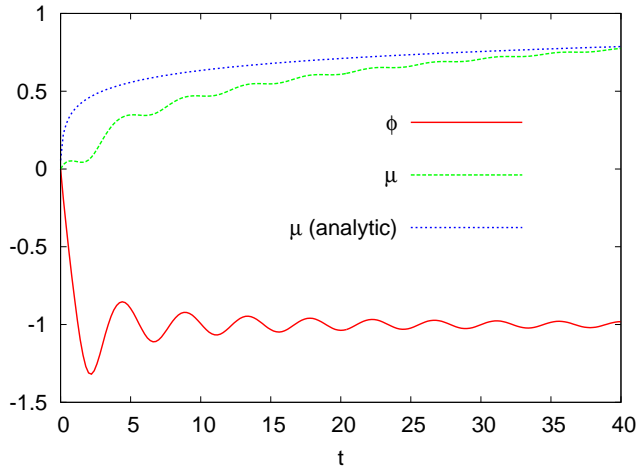}}
\caption{Typical evolutions of dilaton and radius for the stabilizing solutions with the same initial conditions as in Fig. \ref{diverge_com} except for $\lambda=-1, \eta=1$. 
The analytic solutions (\ref{stab_in d-w}) are also plotted as a reference. 
The dilaton goes to the VEV ($\phi_V=-\eta$) as oscillating. The radius expands monotonically.
}
\label{stabilize_com_dw}
\end{center}
\end{figure}

Similar to $\lambda>0$, the potential does not work for the large initial velocity of the dilaton, $\dot{\phi}_0>3\dot{\mu}_0$ and $\dot{\phi}_0>6\dot{\mu}_0$. 
As discussed before, these behaviors are not suitable for our expected situation, and so we examine the other initial conditions where the potential works. 
In such case, the new evolution comes out: the radius grows monotonically, while the dilaton asymptotes to the VEVs ($\phi_V=\pm \eta$), settling down to the potential minimum (Fig. \ref{stabilize_com_dw}). 
The critical velocity $\dot{\phi}_C$ determines the VEV ($\phi_V=-\eta$ or $\eta$), where the dilaton is stabilized in the end.
For the initial velocity of
$\dot{\phi}_0<\dot{\phi}_C$, the radius monotonically grows and the dilaton asymptotically goes to the negative VEV $\phi\to -\eta$, while for $\dot{\phi}_0>\dot{\phi}_C$, the dilaton asymptotically goes to the positive VEV $\phi\to \eta$. 
The former case corresponds to a stabilizing dilaton at the weak coupling. 
In Fig. \ref{stabilize_com_dw}, we plot the numerical results and the analytic solution 
Eq. (\ref{stab_in d-w}) with appropriate integration constants, $\mu_*,\dot{\mu}_*$. 
Note that the forbidden region $E\lsim0$ exists near the critical velocity 
$\dot{\phi}_C$ in the present case (Fig. \ref{fig:v=0_sol}) and then that the allowed region of $\dot{\phi}_0$ which makes the dilaton go to the positive VEV ($\phi_V=\eta$) is narrow since the larger value of $\dot{\phi}_0$ yields the ``growing solution'' indicated in the $V\simeq0$ case of Fig. \ref{fig:v=0_sol}. 
Therefore, the dilaton goes to the negative VEV for a very wide range of parameters. In the initially expanding case, for any value of $\dot{\phi}_0<0$, the radius monotonically grows and the dilaton asymptotes to the negative VEV because $\dot{\phi}_C$ is positive for $H_0>0$.  
The result of these behavior does not depend on $\eta$.


\subsubsection{Effective potential picture}
\label{subsubsec:effective picture modelII}

At first sight, the time evolution of the dilaton under the double-well potential may seem to contradict our intuition. 
As mentioned above, however, the effective potential of the dilaton in the Einstein frame is $W\simeq-\lambda((\phi/\eta)^2-1)^2e^{\phi/2}$ from 
Eq. (\ref{eq:effective potential}). 
Therefore, for $\lambda>0$, there exist two valleys on both sides of the potential $|\phi|\to \infty$, and the slope of the valley existing in the strong coupling regime 
($\phi>0$) is  steeper than in the weak coupling regime by a factor $e^{\phi/2}$. 
Notice that the dilaton falling down to the positive side of the potential corresponds to the rapidly expanding radius, while the dilaton falling down to the negative side corresponds to the rapidly contracting radius. 
Hence when the dilaton starts from the origin, it is easier to make the dilaton roll into the positive side and fall down. 
Then the radius grows rapidly for a very wide range of initial velocities $\dot{\phi}_0$. 
This observation is consistent with the above numerical results.

As for $\lambda<0$, it is basically equivalent to the usual double-well potential picture in terms of $W$ except for its different slope by a factor $e^{\phi/2}$, where the dilaton rolls toward the VEV. 
In this picture, when the dilaton starts from the origin with $\phi_0\lsim0$ and $\dot{\phi}_0\lsim0$, the dilaton always goes to the negative VEV, which is again consistent with the above numerical results.

\subsection{Evolution of large and small radii
\label{sec: IV C}}
Similar to the exponential potential case, we finally discuss evolutions with different radii in the double-well potential case.

For $\lambda>0$, as a result, each radius evolves in the same way: the contracting solution and rapidly growing one described as in Fig. \ref{each_dim2_ex}. 
We cannot find a solution in which each radius evolves differently. 
The asymptotic behavior of the radius is determined by the behavior of the dilaton, i.e., rolling down to the positive side of the valley or rolling down to the negative side of the valley in the effective potential picture, which corresponds to the rapidly growing radius and the contracting one, respectively. 
Therefore, both radii evolve in the same manner since the potential acts on each radius in the same way, which is clearly seen in Eqs. (\ref{Basic_eq1_dilaton}). 
For $\dot{\phi}_0<0$ and  $|\dot{\mu}_0|, |\dot{\nu}_0|\lsim O(1)$, both radii always become the rapidly growing solutions as long as $\eta\gg1$.

\begin{figure}[htmp]
\begin{center}
\scalebox{0.36}{\includegraphics[angle=-90]{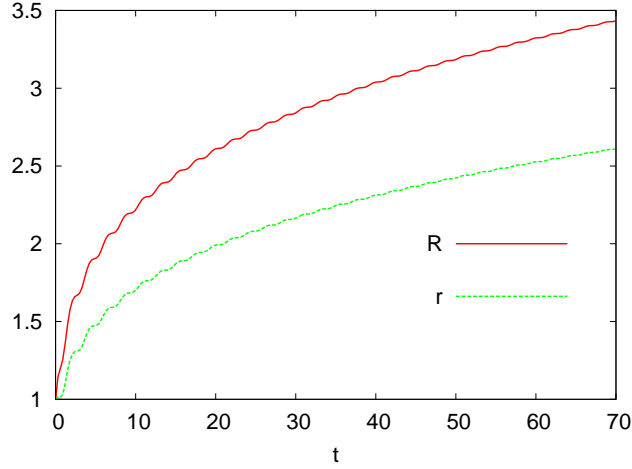}}
\caption{Time evolution of radii large spatial $d=3$ 
dimensions ($R=e^{\mu}$) and small 
$6$ dimensions ($r=e^{\nu}$) for the typical stabilized 
dilaton solutions. The initial conditions are the 
same as in Fig. \ref{each_dim2_ex} except for $\lambda=-1$ and $\eta=0.5$.}
\label{each_dim1_dw}
\end{center}  
\end{figure}
As for $\lambda<0$, similar to the above case of $\lambda>0$, both radii evolve in a same manner: the contracting, growing, and stabilized dilaton solutions. 
We plot the typical evolution of stabilized dilaton solution in Fig. \ref{each_dim1_dw}. 
For $\dot{\phi}_0<0$ and 
$0<\dot{\mu}_0, \dot{\nu}_0\lsim O(1)$, both radii always become 
stabilized dilaton solutions, and such behavior does not depend on $\eta$.

\section{Thermal equilibrium of String Gas 
\label{sec:thermal equilibrium}
}

In order to resolve the dimensionality problem, the annihilation of winding strings play a critical role in the context of the BV scenario. 
The interaction rate $\Gamma$ of annihilation, equivalently, the process where any winding/anti-winding string pair annihilates to the momentum string pair with the coupling given by $e^{\phi}$, is roughly \cite{Danos:2004jz} 
\begin{align}
\Gamma\simeq 100 \ln \,E \,e^{4\phi}\,,
\label{eq:interaction of annihilation}
\end{align}
where $100$ is a numerical factor of sum over both spins and  momentum states and $E$ is the total energy of string gas. 
The thermal equilibrium condition is given by 
\begin{eqnarray}
     \Gamma  > H ,
     \label{eq:thermal equilibrium}
\end{eqnarray}
where $H$ is the Hubble expansion rate. 
The assumption of the thermal equilibrium of string gas is necessary for the BV scenario to work. 
However, a typical cosmological evolution around the Hagedorn temperature is that the radii of the universe asymptote to a constant value $R_{\infty}$ as described by Eq. (\ref{asympotic_const}). 
As discussed in 
Sec. \ref{secIII.A}, $R_{\infty}$ does not become large without fine-tuning. 
It means that all dimensions are still small and it is inconsistent with our large four dimensions. 
Nevertheless, if the radii (of three dimensions) asymptote to a sufficiently large value, the universe could be matched to a radiation dominated phase, and only such a case is a viable scenario. 
Therefore, we require that the asymptotic radius will be larger than a critical radius $\bar{R}$, i.e., 
$R_{\infty}>\bar{R}$. 
Here $\bar{R}$ is a characteristic scale which divides a 
``small'' radius from a ``large" one (Sec. \ref{Hagedorn_regime}).
Then, from (\ref{eq:eoms1}) and (\ref{asympotic_const}), we obtain a constraint equation on the initial value of dilaton as

\begin{align}
e^{\psi_0}<{9{\dot{\mu}_0}^2\over E_0}\biggl
[\Bigl({(\bar{R}e^{-\mu_0})^{{3}}+1\over (\bar{R}e^{-\mu_0})
^{{3}}-1}\Bigr)^2-1\biggr]\,, 
\label{eq:dilaton constraint}
\end{align}
where the subscript $0$ implies the value at an initial time. 
Substituting this initial constraint into 
(\ref{eq:interaction of annihilation}), the interaction rate at an initial state is roughly estimated as 

\begin{eqnarray}
  \Gamma ~\lesssim 100 \ln E \Bigl({9\dot{H}_0^2\over E_0}\Bigr)^2
  O(10^{-1})\lesssim \Bigl(O(10^{-1}) H_0^3\Bigr)H_0\,,
\end{eqnarray}
where we have taken $\bar{R}\sim 3$, 
$\mu_0\sim 0$ and sufficiently high energy, $E\sim 100$. 
This shows that the thermal equilibrium condition is not satisfied at an initial condition as long as the adiabatic condition 
$H_0 \lesssim 1$ holds. 
In what follows, we test the thermal equilibrium condition 
(\ref{eq:thermal equilibrium}) for our models.

\begin{figure}[thmp]
\begin{center}
\scalebox{0.36}{\includegraphics[angle=-90]{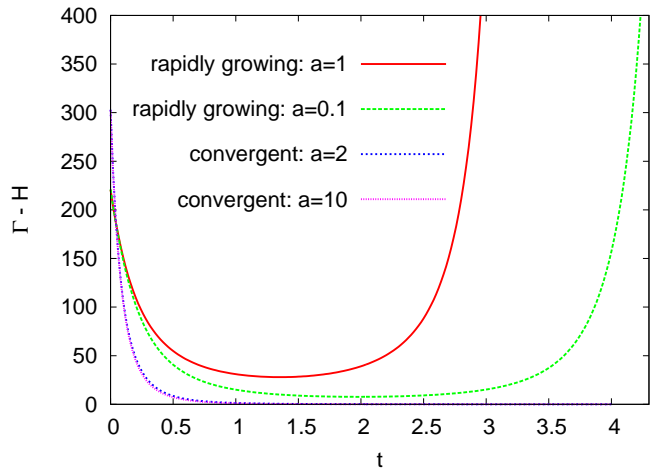}}
\caption{
Plot of $\Gamma-H$ for the model of exponential potential ($b=1$). 
The initial conditions are $\mu_0=0, \phi_0=-0.1,\dot{\mu}_0=0.3$ and 
$\dot{\phi}_0=-1$ (the rapidly growing solutions as red and green lines), or 
$\dot{\phi}_0=-3$ (the convergent solutions as blue and pink lines). 
$\Gamma-H>0$ means thermal equilibrium. 
For the rapidly growing solutions, the weak coupling condition ($e^{\phi} < 1)$ breaks down at $t\gtrsim 3$ for $a=1$ and $t\gtrsim 4$ for 
$a=0.1$.}
\label{thermal_ex}
\end{center}  
\end{figure}

\subsection{Exponential potential case}

\subsubsection{$b>0$ case}

The exponential potential with $b>0$ and $a>1$ allows two types of trajectory for its solutions, i.e., the rapidly growing and convergent ones, described by the 
analytic solutions (\ref{converge_sol}) and 
(\ref{diverge_sol}), respectively. 
Firstly, the convergent solution makes the radius grow toward the asymptotic value of Eq. (\ref{asympotic_const}), yielding the constraint of Eq. (\ref{eq:dilaton constraint}). 
However, once the potential term is included, the constraint (\ref{eq:dilaton constraint}) does not directly restrict the initial value of dilaton at $t=t_0$, but it provides a constraint at a late time, $t=t_*$. 
Therefore, it could relax the constraint on the initial dilaton value, resulting in better realization of the thermal equilibrium. 
Figure \ref{thermal_ex} plots the thermal equilibrium condition with time: $\Gamma-H>0$ is equivalent to the thermal equilibrium condition. 
 From this figure, we find the thermal equilibrium is realized until 
$t\sim 1$, which is equivalent to the time scale on which the potential term works, $t\lsim t_*$. 
This duration does not depend sensitively on $a$. 
The final scale of the radius is, however, not so large, and it may be insufficient to continue the decompactification process. 
Nevertheless, a noteworthy point is that the thermal equilibrium is realized at the initial phase, contrary to the naive scenario.

Secondly, the rapidly growing solution is the solution in which the radius grows unboundedly, and thus the initial constraint cannot be applied in this case.
Contrary to the convergent case, the dilaton grows toward the strong coupling, and the weak coupling condition breaks down. 
We see that the thermal equilibrium is also realized as seen in Fig. \ref{thermal_ex}. 
So the thermal equilibrium is at least realized until the time of the violating, even though the time of the breaking becomes longer as 
$a$ decreases. The duration of thermal equilibrium becomes large at most by a factor of two. 
For $a\leq 1$, all solutions are the rapidly growing type, and, for example, the duration is about $t\sim 5$ for $a=0.01$.

\subsubsection{$b<0$ case} 

In this case, all solutions with $a>1$ that satisfy the weak coupling condition become convergent solutions. 
Thermal equilibrium is the same as the convergent solutions in $b>0$, and hence the initial equilibrium continues until $t\sim O(1)$.

\begin{figure}[thmp]
\begin{center}
\scalebox{0.36}{\includegraphics[angle=-90]{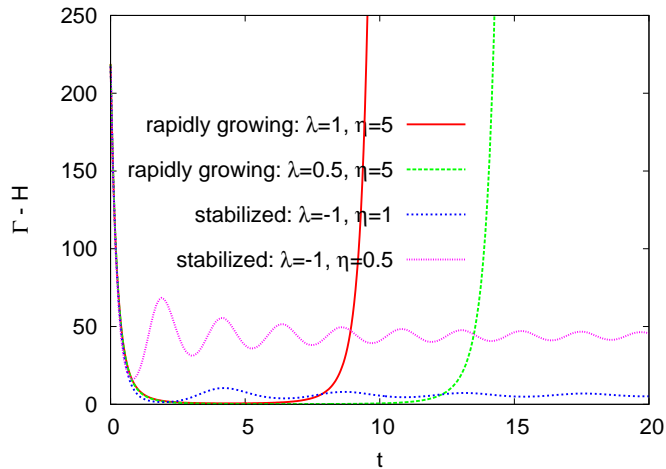}}
\caption{Plot of $\Gamma-H$ for the model of double-well potential. 
The initial conditions are 
$\mu_0=0, \phi_0=-0.1,\dot{\mu}_0=0.3$ and $\dot{\phi}_0=-1$.
For the rapidly growing solutions, the weak coupling condition breaks down at $t\gtrsim 10$ for $\lambda=1$ and $t\gtrsim 15$ for $\lambda=0.5$.}
\label{thermal_dw}
\end{center}  
\end{figure}

\subsection{Double-well potential case}
\subsubsection{$\lambda>0$ case}

All solutions with $\eta\gg 1$ that satisfy the weak coupling condition become rapidly growing solutions. 
In Fig. \ref{thermal_dw} we plot the thermal equilibrium condition. 
The basic behavior of thermal equilibrium is the same as in the growing solution for the exponential potential with 
$b>0$, except for the time scale: the initial equilibrium is realized until $t\sim O(10)$ where the weak coupling condition violates. 
The thermal equilibrium is only marginally satisfied at almost all times until the violating.

\subsubsection{$\lambda<0$ case}

In this case, all solutions which satisfy the weak coupling condition become the stabilized dilaton solutions. 
If we take $\eta$ to be small, e.g., $\eta \lesssim O(1)$, the thermal equilibrium continues unboundedly (Fig. \ref{thermal_dw}). 
This is because the solution stabilizes the dilaton at the VEV ($\phi = -\eta$) and then the interaction rate $\Gamma\propto e^{4\phi}$ asymptotes to $e^{-4\eta}\sim O(0.1)$. 
At the same time, the Hubble expansion rate asymptotes to zero, $H\propto 1/t\to 0$, and then we have 
$(\Gamma-H)|_{t\to\infty}>0$. As Fig. \ref{thermal_dw} shows, the thermal equilibrium continues unboundedly as long as $\eta\lsim 1$. 
This result is the best situation for the BV scenario in our models. 
As for $\eta> 1$, the thermal equilibrium is only marginally satisfied at late times, $\Gamma-H \approx 0$.

\section{summary and discussion}

We have studied the thermal equilibrium of string gas in the Hagedorn regime where the universe is in high energy. 
Thermal equilibrium is the one of the important assumptions for the BV scenario in order to induce the dynamical decompactification of three large spatial dimensions. However, the initial thermal equilibrium condition of string gas is not realized in the original scenario based on dilaton-gravity. 
To resolve this difficulty, we have explored possibilities for avoiding the issue. 
As a first step to tackle this problem, we have studied a minimal modification of the original model, by introducing a potential term of the dilaton. This simple setup allows us to study the system rather extensively. 
However, this does not mean that stabilization of the dilaton or the 
effects of the potential term is a necessary ingredient. 
We wish to emphasize that effects of matter 
(e.g., flux or any kind of corrections, etc.) would not be negligible, and taking into account such effects, we could avoid the issue of thermal equilibrium in the early universe. 
We expect our simple setup will provide implications for such effects.

We have taken the dilaton potential to comprise two simple 
potentials, i.e., the exponential potential and the double-well 
potential, and have analyzed both the dynamics of the system and 
the thermal equilibrium condition at the initial 
stage of the universe. 
Even though we have mainly studied the evolution of the scale factor 
with same radii, there is no significant difference 
in the typical evolutions of different radii. 
Based on the solutions, we have examined whether they satisfy three basic assumptions, i.e., the adiabatic condition $H_0\lsim O(1)$, 
the weak coupling condition and the thermal equilibrium condition $\Gamma>H$. 
As a result, we find the following cases.
\begin{description}
	\item[Exponential potential : $V=b\,e^{2a\phi}$] 
\end{description}
{\it (i) $b>0$ :  the convergent and rapidly growing solution}
\\[.5em] 
The convergent and the rapidly growing solutions for $a>1$ are acceptable cases. 
The former case is that the radius converges to a constant value and the dilaton rolls monotonically to the weak coupling regime. In the latter case, both radius and dilaton asymptotically grow in short time. 
These different evolution is determined by whether the initial velocity of dilaton $\dot{\phi}_0$ is smaller or larger than the critical velocity $\dot{\phi}_C$. 
On the other hand, for $a\le 1$, all numerical solutions are asymptotically the rapidly growing solutions. 
Both of two solutions satisfy the thermal equilibrium condition during some initial time. 
\\[.5em] 
{\it (ii) $b<0$ :  the convergent solution} 
\\[.5em]
In this case, the convergent solutions are sensible solutions since the others are contracting evolutions. 
For any value of $a$, all numerical solutions satisfy the thermal equilibrium condition during the initial time.

\begin{description}
	\item[Double-well potential : 
	$V={\lambda\over 4}(({\phi\over \eta})^2-1)^2$] 
\end{description}
{\it (i) $\lambda>0$ : the rapidly growing solution}
\\[.5em] 
For $\eta\gg 1$, all numerical solutions are rapidly growing solutions, and they satisfy the thermal equilibrium condition during some initial time which is relatively longer than the above cases. 
\\[.5em]
{\it (ii) $\lambda<0$ :  the stabilized dilaton}
\\[.5em]
In this case, the dilaton is stabilized and it does hold the weak coupling condition. 
For any value of $\eta$, all numerical solutions are reduced to the stabilized dilaton solutions. If we choose $\eta\lsim O(1)$, the thermal equilibrium continues unboundedly.

From these results, we conclude that it is possible to realize the thermal equilibrium of string gas at the initial Hagedorn regime. 
At the end of the paper, we would like to ask if the evolutions can be matched to a late-time universe. 
For the convergent solutions, the universe will not enter into the large-radius phase  $R>\bar{R}$, because $\bar{R}$ does not become large 
without fine-tuning, as discussed above. 
Besides, the time scale of the thermal equilibrium will too short 
for the BV mechanism to work, compared with Hubble time. 
Similarly, for the rapidly growing solutions, the short time scale will be a problem in the case of the exponential potential. 
However, in the double-well potential case, the duration becomes relatively longer and the situation is better.
Nevertheless, it remains as a problem that the dilaton asymptotes to the strong coupling, where we cannot predict what will happen.

On the other hand, 
the stabilized dilaton solutions can make the radius grow as $R\propto t$, and it would be possible to match it to a late-time universe. 
Moreover, the thermal equilibrium continues unboundedly. 
This solution is the best case among all examples presented in our paper. 
It implies that the (quasi-) stabilized dilaton at an early stage of the universe improves the situation. Besides, this example implies an interesting possibility for constructing a model that resolves the stabilization and dimensionality problem at the same time.
It remains an interesting question whether we can build a model which can be embedded in string theory resolving all problems described above.

\acknowledgments
We would like to thank Kei-ichi Maeda and 
Shun Saito for their comments and/or discussions on this work. 
The work of YT and HK are supported by JSPS and YT is also supported by 
Grants-in-Aid for the 21th century COE program ``Holistic Research and 
Education Center for Physics of Self-organizing Systems" 
at Waseda University.

\appendix
\section{Effective potential}
\label{effective_potential} 

In this Appendix we will show the effective potential of the dilaton in the Einstein frame derived by conformal transformation. Quantities in the conformal metric will be denoted by a tilde. 
We will denote the conformal factor by $\Omega^2$ which is a function of the $D$ dimensional spacetime coordinates $x^{\mu}$. 
The conformally transformed metric is 
$
\til{g}_{\mu\nu}=\Omega^2 g_{\mu\nu}\,,
$
and the determinant of the metric scales as 
$
\sqrt{-\til{g}}=\Omega^D\sqrt{-g}\,.
$

Consider the dilaton-gravity action in 
$D$ dimensions, 
\begin{eqnarray}
  S 
  =
  \int d^{D} x \sqrt{|g|}e^{-2\phi} \Bigl[ R+4(\nabla \phi)^2 + V \Bigr]\,.
\end{eqnarray}
Under the conformal transformation, the action becomes \cite{Lidsey:1999mc}
\begin{eqnarray}
  S=\int d^{D} x 
  \sqrt{|\til{g}|} e^{-2\phi} 
  \Omega^{2-D}  \Bigl[ \til{R} 
  +4(\til{\nabla} \phi)^2 
  +(D-2)(D-1)\Omega^{-2}(\til{\nabla}\Omega)^2
  +\Omega^{-2}V 
  +4(D-1)\Omega^{-1}(\til{\nabla}\Omega)(\til{\nabla}\phi)\Bigr]\,.
\end{eqnarray}
The last term comes from integrations by parts. 
If we set the conformal factor as 
$
\Omega^{2-D}=e^{2\phi}\,,
$
the action in the Einstein frame is 
\begin{eqnarray}
  S=\int d^{D} x 
  \sqrt{|\til{g}|} \Biggl[ \til{R} 
  +4\Bigl(1-{D-1\over D-2}\Bigr)(\til{\nabla} \phi)^2 
  +e^{4\phi\over D-2 }V \Biggr]
\,. 
\end{eqnarray}
In the case $D=10$ the Lagrangian of the dilaton which is minimally coupled to the metric in this frame, is reduced to 
\begin{eqnarray}
L_{\phi}=-{1\over 2}(\til{\nabla}\phi)^2-W(\phi)\,,
\end{eqnarray}
with the canonically normalized potential 
\begin{align}
 W(\phi) = -e^{\phi\over 2}V(\phi)\,.
\end{align}

\section{Autonomous phase plane}
\label{phase_plane}

\begin{figure}[htmp]
\begin{center}
\includegraphics[width=8cm]{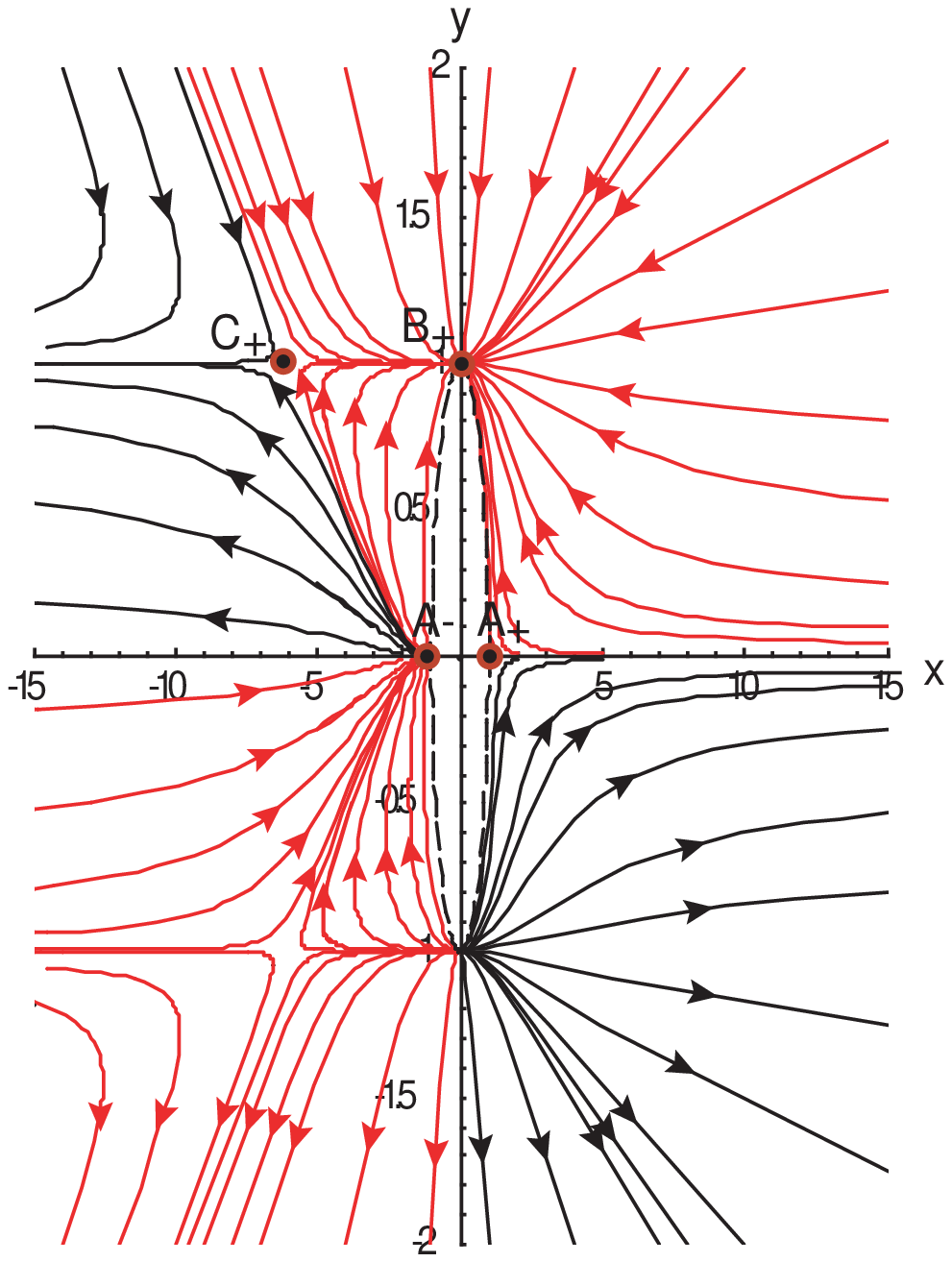}
\includegraphics[width=8cm]{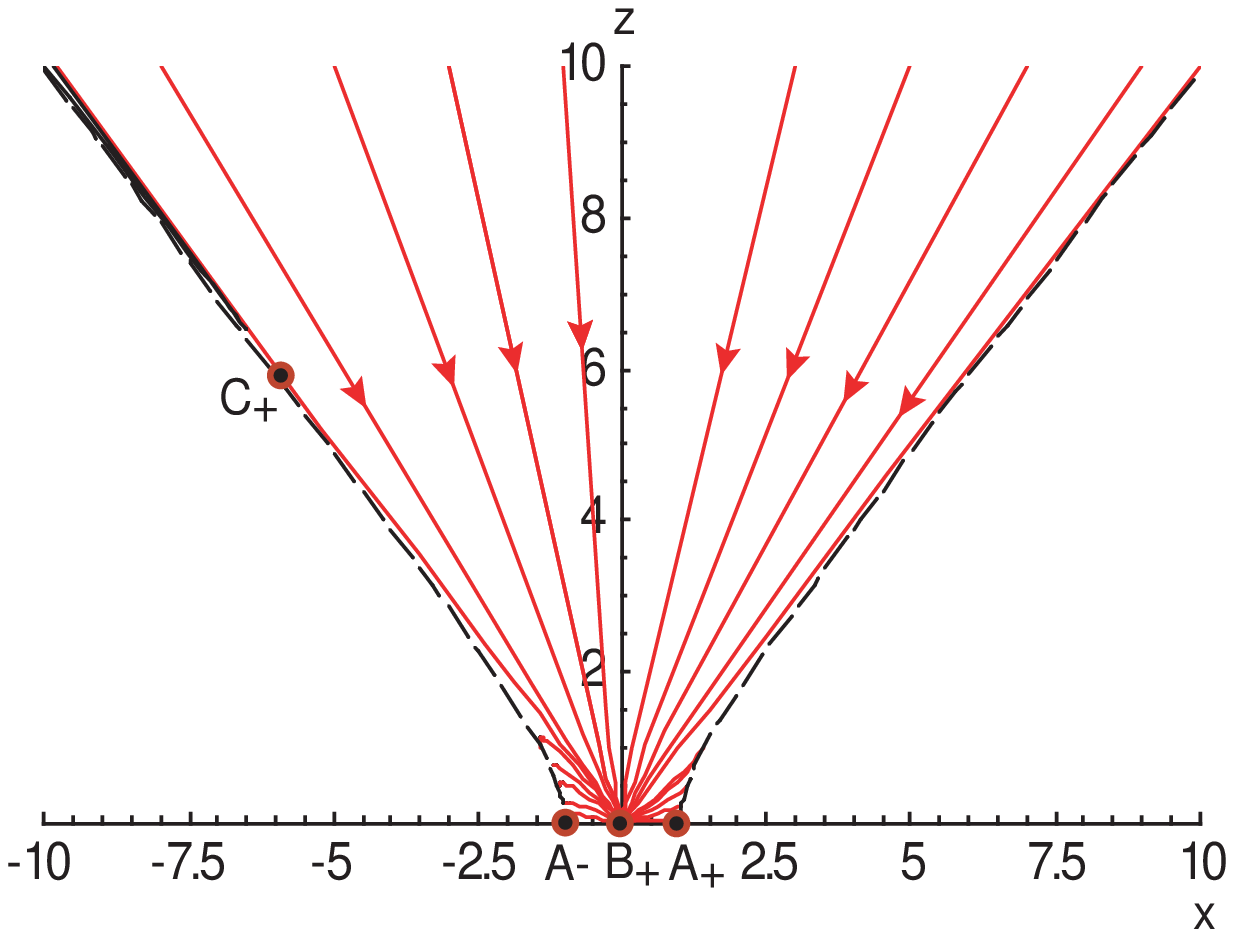}
\caption{
The phase trajectories for $a=2$ in $(x,y)$ and $(x,z)$ planes.
There are four fixed points in the upper half-planes. 
$B_{+}$ is the attractor point for the rapidly growing solution. 
$C_{+}$ divides the convergent solution from the rapidly growing one. 
The red lines show the trajectories approaching the rapidly growing solutions, and the black lines show the trajectories approaching the convergent solutions. 
The arrows represent the direction of time evolution.
The color of the lines indicates the end state of these solutions. 
An enlarged figure showing the behavior of the black lines in $(x,z)$ plane is depicted in Fig. \ref{a=2_large}. 
The dotted lines represent the constraint $1=x^2+y^2$ or $1=x^2-z^2$, which means the forbidden region in the phase space.
}
\label{a=2_xy}
\end{center}  
\end{figure}

\begin{figure}[ht]
\begin{center}
\includegraphics[width=8cm]{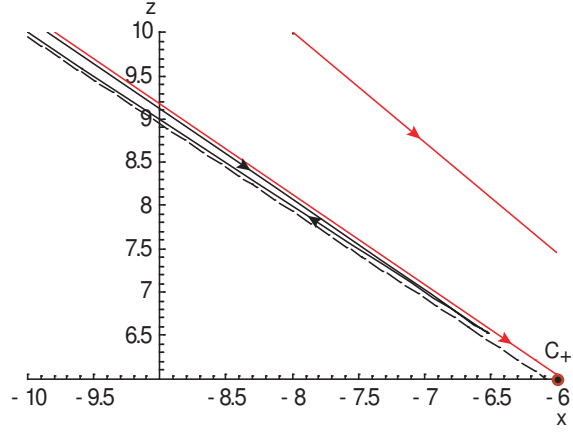}
\caption{Enlarged figure near the fixed point $C_{+}$ in 
Fig. \ref{a=2_xy}. 
The trajectory of the black line approaches $C_{+}$ and returns to $z\to \infty$.}
\label{a=2_large}
\end{center}  
\end{figure}

\begin{figure}[htmp]
\begin{center}
\includegraphics[width=8cm]{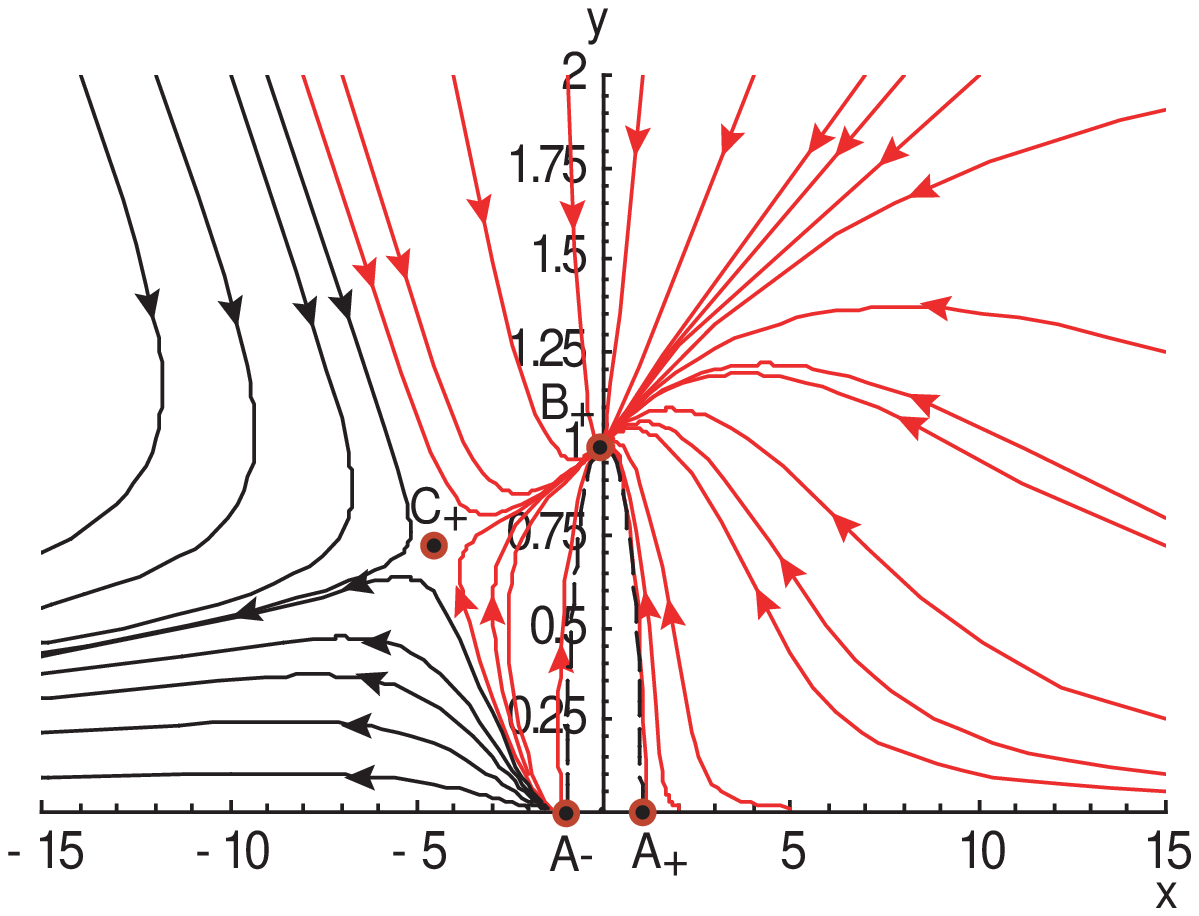}
\includegraphics[width=8cm]{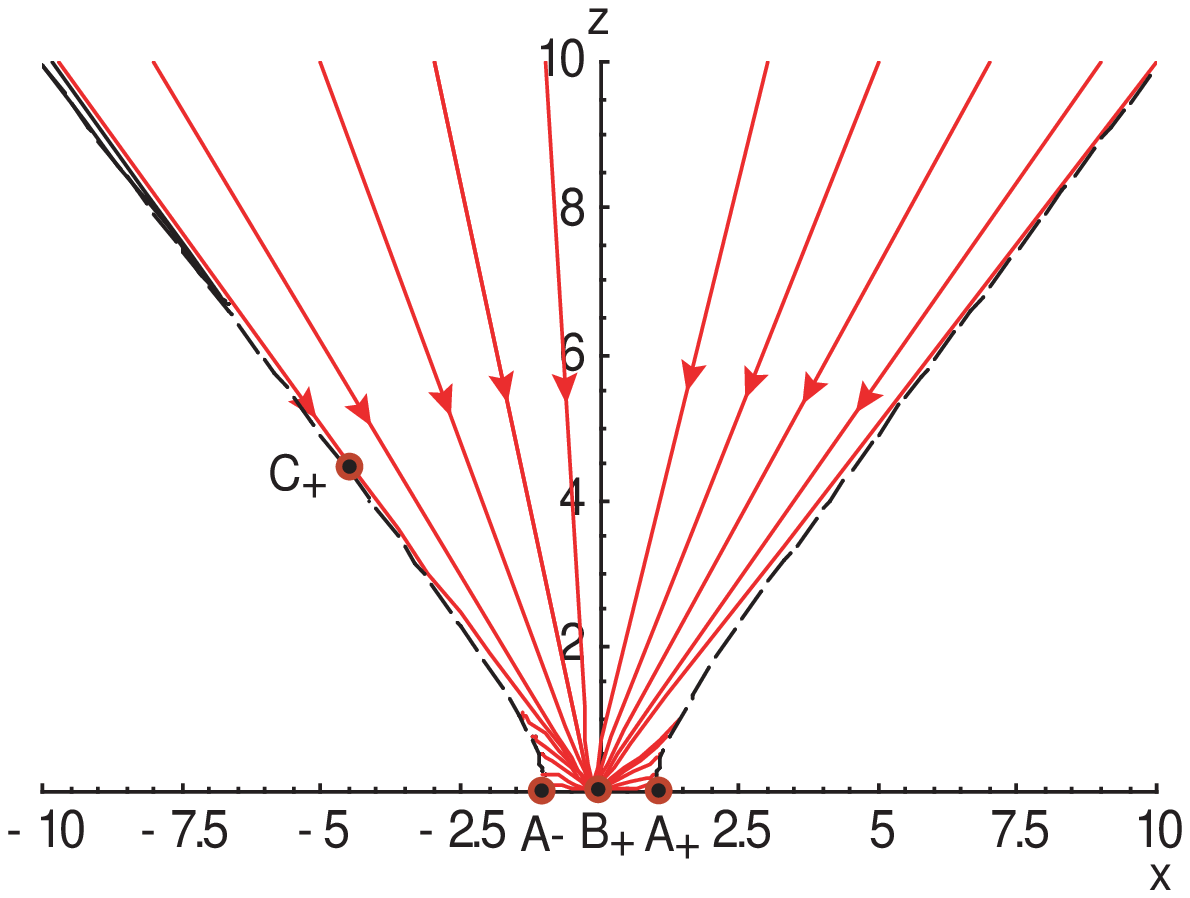}
\end{center}
\caption{
The phase trajectories for $a=3$ in $(x,y)$ and $(x,z)$ planes.
The four fixed points are the same as in Fig. \ref{a=2_xy}.}
\label{a=3_xy}
\end{figure}

\begin{figure}[htmp]
\begin{center}
\includegraphics[width=8cm]{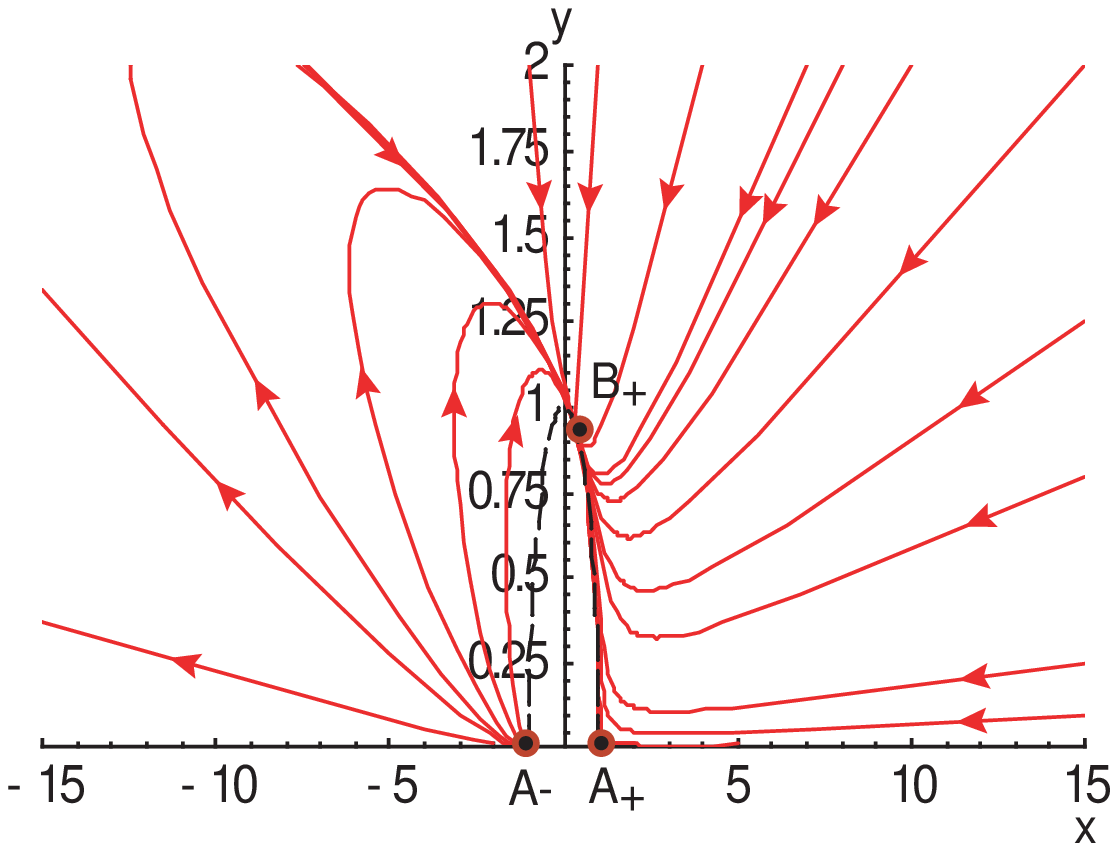}
\includegraphics[width=8cm]{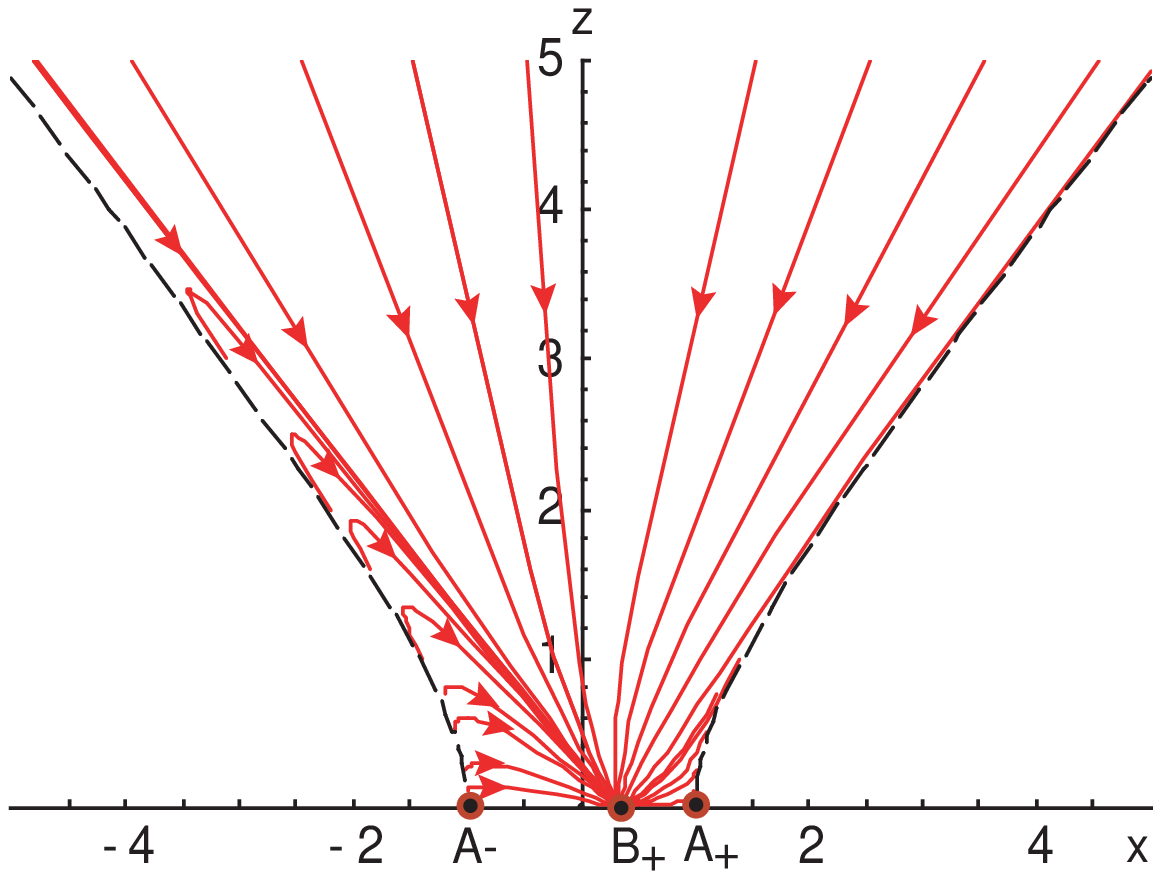}
\end{center}
\caption{
The phase trajectories for $a=1$ in $(x,y)$ and $(x,z)$ planes.
$A_+$, $A_-$ and $B_+$ are the fixed points.
$C_{+}$ does not exist, contrary to the other cases, and hence there is no trajectory approaching convergent solutions. 
All trajectories approach $B_{+}$, and all these solutions are asymptotically rapidly growing solutions.}
\label{a=1_xy}
\end{figure}

\begin{figure}[htmp]
\begin{center}
\includegraphics[width=8cm]{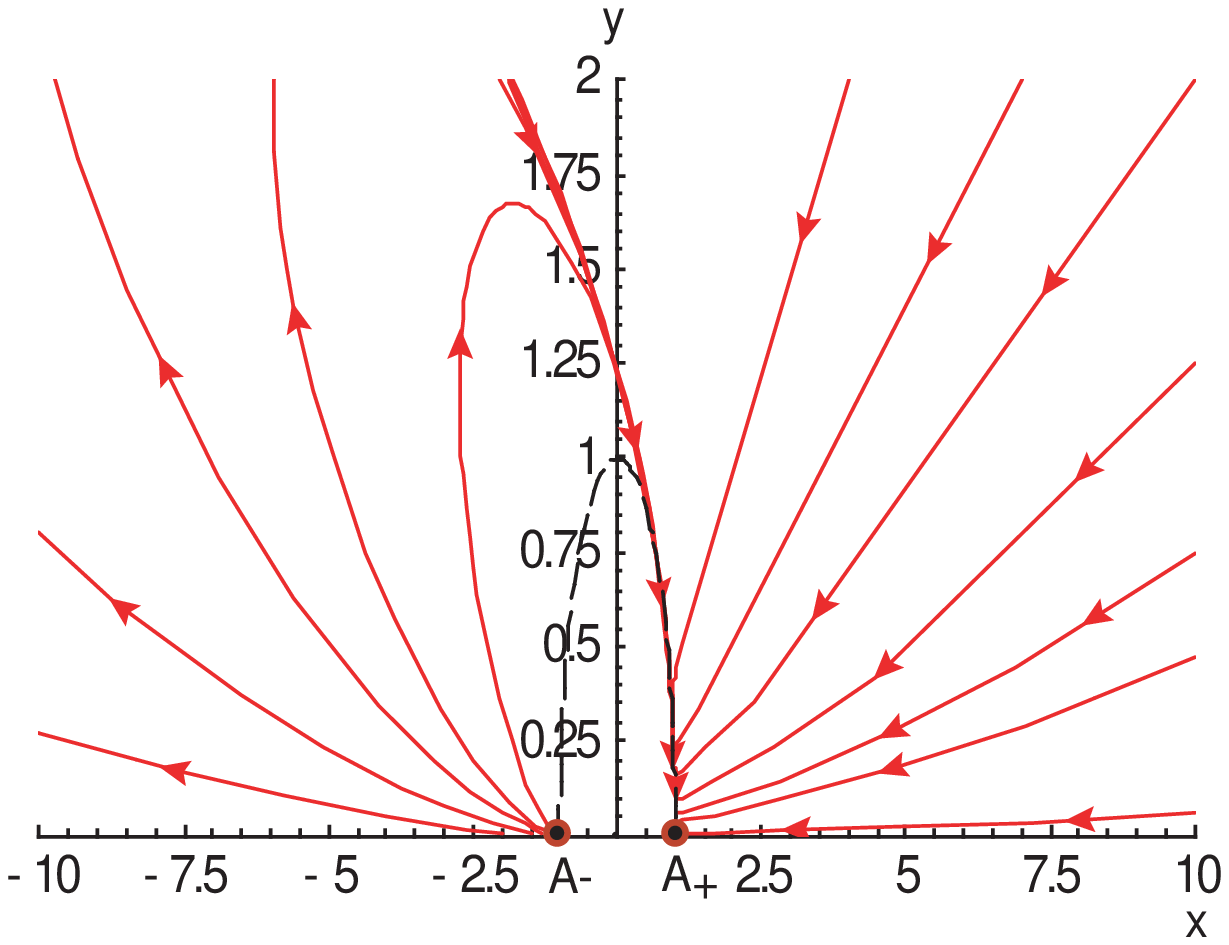}
\includegraphics[width=8cm]{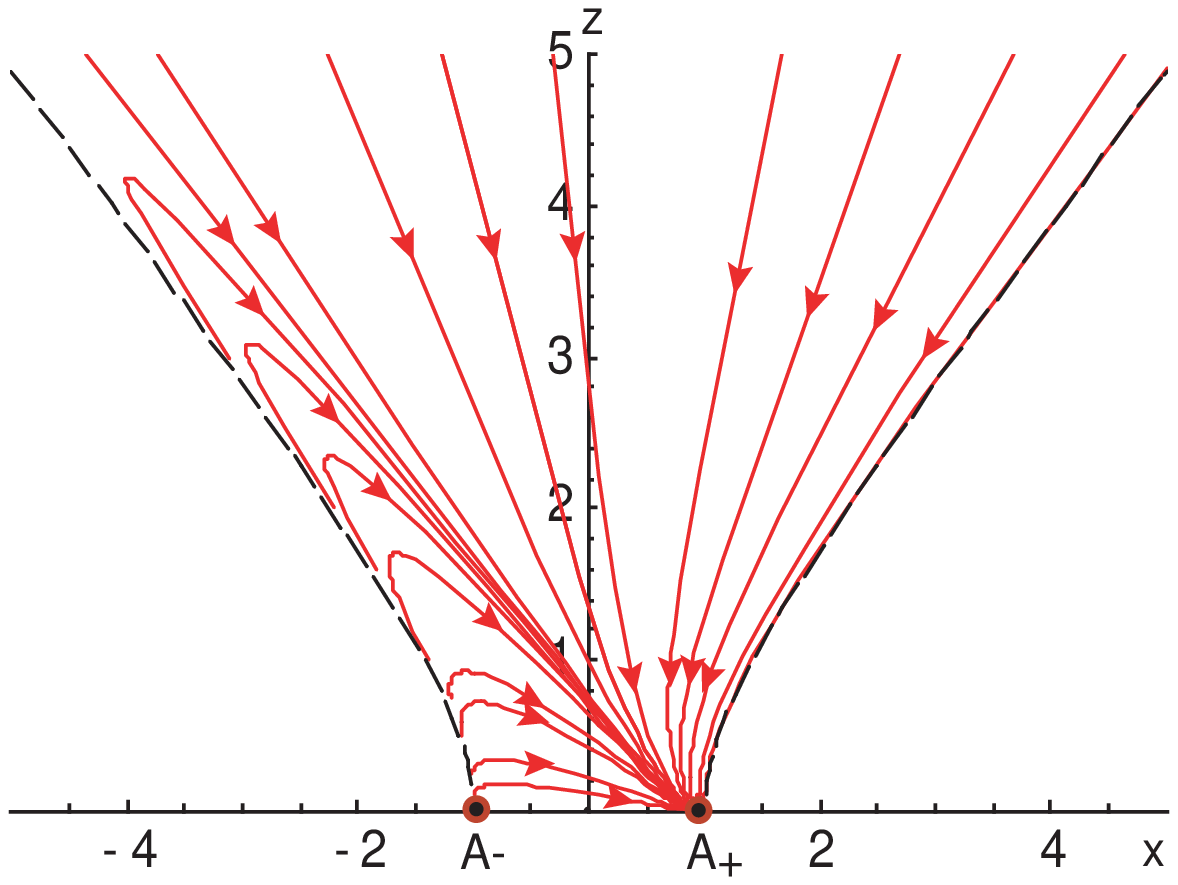}
\end{center}
\caption{
The phase trajectories for $a=1/2$ in $(x,y)$ and $(x,z)$ planes.
$A_+$, $A_-$ are the fixed points.
$B_{+}$ and $C_{+}$ do not exist.
$A_{+}$ is the attractor point replaced by $B_{+}$ corresponding to the rapidly growing solutions. 
All trajectories approach $A_{+}$, and these solutions are asymptotically the rapidly growing solutions.}
\label{a=0.5_xy}
\end{figure}

In this Appendix, we describe the asymptotic behavior of the system with the exponential potential.  
We define the dimensionless phase-space variables 
\cite{Copeland_L_W, Allen_Wands}
\begin{align}
 x \equiv {2\dot{\phi}-9H \over 3H},~~
 y \equiv {\sqrt{|V|}\over 3H},~~
 z \equiv {\sqrt{\rho}e^{\phi}\over 3H}\,.
\end{align}
The Eqs. (\ref{eq:simple form of EOM mu, phi}) are reduced to a so-called plane autonomous system, which consists of three evolution equations and a constraint equation, 
\begin{eqnarray}
&&x'={3\over2}\bigl[1-x^2+(1-a)y^2-3axy^2\bigr]\,,\cr
&&y'=-{3\over2}y\bigl[(2-a)x+3ay^2-3a \bigr]\,,\cr
&&z'=-{3\over2}z(x+3ay^2)\,,
\cr
&&1=x^2+y^2-z^2\,.
\end{eqnarray}
Here a prime denotes a derivative with respect to the number of $e$-foldings, $\ln(R)=\mu$. 
As the system is symmetric under the reflection $y \to -y$ and $z \to -z$, we mainly consider only the upper half-planes, $y\ge0$ and $z\ge0$, in the following discussion
\footnote{
$y<0$ corresponds to the contracting evolution ($\dot{\mu}<0$), and the lines of phase-trajectory are the same as $y\ge0$, except the direction of the arrow (Fig. \ref{a=2_xy}). 
}.
We solve this system numerically to find the phase-trajectories. 
In Figs. \ref{a=2_xy} - \ref{a=0.5_xy}, we show phase trajectories on the $y$-$x$ and $z$-$x$ planes for $a=1/2,~1,~2,~3$. 
In this system, there are three types of fixed point (critical point), defined by $x'=0,y'=0,z'=0$: 
\begin{eqnarray}
A_{\pm} & :& x=\pm1,\quad y=0,\quad z=0\,, 
\cr
B_{\pm} &:& x={2-a\over3a},
   \quad y=\pm{2\sqrt{2a^2+a-1}\over 3a}, 
   \quad z=0 
\,, 
\cr
C_{\pm}&:& 
   x=-{3a\over a-1},
   \quad y=\pm{1\over\sqrt{a-1}}, 
   \quad z=\pm{\sqrt{8a^2+3a-2}\over a-1}\,.
\end{eqnarray}
$C_{\pm}$ is vanishing for $0<a\leq1$, and 
$B_{\pm}$ is also vanishing for $0< a\leq 1/2$. 
For $a\leq1$, there are four fixed points $A_+$, $A_-$, $B_{+}$ and $C_{+}$ in the regime $y\ge 0$ and $z\ge 0$. 
For $1/2<a\leq 1$, there are three fixed points, $A_+$, $A_-$ and $B_{+}$.
For $0<a\leq 1/2$,  $A_+$ and $A_-$ are the fixed points.
It can be understood from the global behaviors on the phase-space that $A_{\pm}$ may correspond to an unstable node (or saddle node).

In the following we show that the critical point $B_{+}$ describes the rapidly growing solution of Eq. (\ref{diverge_sol}) and 
$C_{+}$ divides the convergent solution 
(\ref{converge_sol}) from the rapidly growing one. 
In order to see the behavior of these solutions on the phase-space, we rewrite these solutions in terms of the phase-space variables ($x,y,z$). 
From Eqs. (\ref{diverge_sol}), 
the rapidly growing solution with $b=\pm1$ at late times ($\dot{\mu}\to \infty$) gives 
\begin{eqnarray}
&& x = {2\dot{\phi}-9\dot{\mu}\over 3\dot{\mu}}\propto
{1\over 3}\,,
\cr
&& y = {e^{a\phi}\over 3 \dot{\mu}}
\propto 1/ \cos \Bigl[a C_1(t-t_*+C_2)\Bigr] \,,\cr
&&
z = {\sqrt{\rho}e^{\phi}\over 3\dot{\mu}}
  = {\sin\Bigl[a C_1(t-t_*+C_2)\Bigr]^{1-1/a}\over 
\cos\Bigl[a C_1(t-t_*+C_2)\Bigr]}\,.
\end{eqnarray}
Therefore, for $a> 1$, $x$ and $y$ asymptote to finite values, while 
$z$ goes to zero at the divergent point $a C_1(t-t_*+C_2)\sim\pi$, as seen in Eq. (\ref{eq:divergent point}). 
The trajectories approach the attractor 
point $B_{+}$, which are plotted by the red lines in Figs.~
\ref{a=2_xy}-\ref{a=3_xy}. 
Similarly, all trajectories for $1/2< a\leq 1$ approach the attractor point $B_{+}$ (Fig.~\ref{a=1_xy}), while for $0< a\leq1/2$ the critical point 
$A_{+}$ plays a role of $B_{+}$ corresponding to the rapidly growing solution (Fig.~\ref{a=0.5_xy}).

On the other hand, the behavior of the convergent solution at late times, when the (shifted) dilaton decreases to the negative infinity, 
$\psi(=2\phi-9\mu) \to -\infty$, follows from Eq. (\ref{converge_sol}): 
\begin{eqnarray}
&& x={\dot{\psi}\over 3\dot{\mu}}\propto 
 - \frac{E_0}{2}(t-t_*)- B \,,
\cr
&& y={e^{a\phi}\over 3 \dot{\mu}}
\propto  e^{(\frac{a}{2}-1)\psi}\,,
\\
&& z={\sqrt{\rho}e^{\phi}\over 3\dot{\mu}}
= e^{- {\psi\over 2}} \,.
\nonumber
\end{eqnarray}
Therefore, $x\to -\infty$ and $z\to \infty$ for $a\ge 2$, while 
$y$ goes to a finite value at late times. 
In Figs.~\ref{a=2_xy}-\ref{a=3_xy}, the trajectories that approach asymptotically the convergent solution are plotted by the black lines.

From these figures for $a>1$, we see two types of asymptotic trajectory: the trajectory approaching $B_{+}$ which represents the rapidly growing solution and the one approaching the convergent solution ($|x|,|z|\to \infty$). 
These two types of trajectory are clearly divided by flows around the fixed point $C_{+}$.
For $0<a\leq1$, all trajectories approach the rapidly growing solution because the fixed point $C_{+}$ disappear as shown in Figs. \ref{a=1_xy} and \ref{a=0.5_xy}.

From these figures, for $a>1$, we can estimate the critical line which divides the rapidly growing solution from the convergent one in the phase-plane $(x,y)$.
The critical line is approximated by the straight line passing through $A_-$ and $C_+$ 
\begin{align}
y\simeq-{\sqrt{a-1}\over 2a+1}(x+1)\,. \quad  (y\ge0)\,.
\end{align}
Therefore, the trajectories that asymptote to 
the convergent solution are given by
\begin{align}
   y~\lsim -{\sqrt{a-1}\over 2a+1}(x+1),
\end{align}
while the trajectories that asymptote to the rapidly growing solution are characterized by the opposite inequality sign. 
This condition can be rewritten in terms of the 
initial dilaton and spacetime variables. Assuming $\phi_0\simeq 0$, the condition for the convergent solution is 
\begin{align}
\dot{\phi_0}~\lsim ~3\dot{\mu_0}-{2a+1\over 2\sqrt{a-1}},
\end{align}
which is applicable for the expanding case $\dot{\mu}_0>0$ ($y>0$).

In our numerical analysis in Sec.~\ref{sec:numerical analysis model I}, 
we vary the initial velocity of the dilaton, fixing other initial conditions. 
From the above equation, the critical velocity for 
the initially expanding case 
that divides the late-time evolutions is approximately given by
\begin{align}
\dot{\varphi}_C\sim 3\dot{\mu}_0-{2a+1\over 2\sqrt{a-1}}\,.
\label{eq:critical velocity}
\end{align}
Any velocity lower than the critical velocity yields the convergent solution, while any velocity beyond the critical one yields the rapidly growing one. 
The analytic estimation (\ref{eq:critical velocity})
is in good agreement with the 
critical velocity $\dot{\phi}_C$ 
obtained from the numerical analysis.
For example,  
we find $\dot{\phi}_C/\dot{\varphi}_C\sim 0.88$ for $\dot{\mu}_0=0.2$  
and $\dot{\phi}_C/\dot{\varphi}_C\sim 1.3$ for $\dot{\mu}_0=0.5$, based on 
our numerical simulation for $a=2$.

For the contracting case $y<0$, the trajectories 
asymptoting to the convergent solutions are obtained from Fig. 
\ref{a=2_xy}: 
\begin{align}
x\gsim 0.  ~~(y<0)\,.
\end{align}
The condition required for the convergent solution is 
$
2\dot{\phi_0}~\lsim ~9\dot{\mu_0}
$,
and then the critical velocity $\dot{\varphi}_C$ for $\dot{\mu}_0<0$ is 
\begin{align}
\dot{\varphi}_C\sim 9/2\dot{\mu_0}.
\label{eq:critical velocity H<0}
\end{align}


\end{document}